\documentclass[12pt]{article}

\usepackage[utf8]{inputenc}      
\usepackage[T1]{fontenc}         
\usepackage{lmodern}             
\usepackage{newunicodechar}
\newunicodechar{ }{~} 
\newunicodechar{⁻}{\textsuperscript{-}} 

\usepackage{amsmath,amssymb,amsthm}

\usepackage{graphicx}
\usepackage{siunitx}             

\usepackage[a4paper,margin=1in]{geometry}

\usepackage{float}               
\usepackage{caption}
\usepackage{booktabs}            

\usepackage[%
  backend=biber,
  style=ieee,
  citestyle=numeric-comp,   
  sorting=none,
  doi=false,isbn=false,url=false
]{biblatex}
\usepackage{hyperref}
\hypersetup{
  colorlinks=true,
  linkcolor=blue,
  citecolor=blue,
  urlcolor=blue,
  pdfpagemode=UseNone
}
\usepackage[capitalize,nameinlink]{cleveref}
\crefname{figure}{Fig.}{Figs.}
\crefformat{figure}{#2Fig.~#1#3}

\usepackage{authblk}

\usepackage{todonotes}

\newcommand{\GaAs}{\ensuremath{\mathrm{GaAs}}}

\newcommand{\AlGaAs}{\ensuremath{\mathrm{Al}_{0.15}\mathrm{Ga}_{0.85}\mathrm{As}}}
\newcommand{\nm}{\nano\metre}
\newcommand{\um}{\micro\metre}
\newcommand{\mm}{\milli\metre}
\newcommand{\kHz}{\kilo\hertz}

\newcommand{\GHz}{\giga\hertz}

\DeclareUnicodeCharacter{0308}{{\"o}}   
\DeclareUnicodeCharacter{2009}{ }       

\title{Compact system development of efficient quantum-entangled photon sources towards deployable and industrial devices}

\author[1]{Yared G. Zena}
\author[1,3]{Moritz Langer}
\author[1]{Ahmad Rahimi}
\author[1]{Abhishikth Dhurjati}
\author[2]{Pavel Ruchka}
\author[2]{Sara Jakovljevic}
\author[1]{Mandira Pal}
\author[4]{Frank H. P. Fitzek}
\author[2]{Harald Giessen}
\author[5]{Juergen Czarske}
\author[3]{Riccardo Bassoli}
\author[3]{Caspar Hopfmann\thanks{Corresponding author, email: caspar\_arndt.hopfmann@tu-dresden.de}}

\affil[1]{Institute for Emerging Electronic Technologies, IFW Dresden, Helmholtzstraße 20, 01069 Dresden, Germany}
\affil[2]{4th Physics Institute and Research Center SCoPE, University of Stuttgart, 70569 Stuttgart, Germany}
\affil[3]{Quantum Communication Networks research group, Deutsche Telekom Chair of Communication Networks, Dresden University of Technology, Germany}
\affil[4]{Deutsche Telekom Chair of Communication Networks, Dresden University of Technology, Germany}
\affil[5]{Chair of Measurement and Sensor Systems Technique, TU Dresden, Dresden, Germany}

\begin{document}

\date{\today}

\maketitle
\begin{abstract}
Entangled photon pair sources are a key enabling technology for quantum communication and networking, yet their deployment beyond laboratory environments is hindered by system-level complexity, limited operational stability, and insufficient industry compatibility. Here, we demonstrate a rack-based, mobile quantum light source architecture based on a semiconductor quantum dot emitter that directly addresses these challenges through modular system integration and automated operation. The source generates polarization-entangled photon pairs with an entanglement negativity 2n of up to \num{0.98 \pm 0.01}, confirming near-maximal entanglement quality. In continuous, hands-off operation over a six-hour time window, the system achieves an average single-photon emission rate of \qty{697\pm 8}{\kHz} and a maximum rate of \qty{740 \pm 7}{\kHz}, while maintaining 2n-value of more than \qty{95}{\percent}. These results are enabled by the integration of optical excitation, collection, cryogenic operation, and control electronics within a standardized rack footprint, together with automated monitoring. By demonstrating simultaneously high entanglement quality, sustained brightness, and long-term operational stability in an industry-aligned system architecture, this work advances semiconductor quantum dot sources toward deployable entangled photon sources for applied quantum photonics.

\end{abstract}

\section{Introduction}
\label{sec:Intro}
Quantum communication is essential for the quantum information exchange between distant communication network nodes, thereby enabling key applications such as physically secure information exchange and distributed quantum computing \cite{Ball2018, Kimble2008}. While early systems focused on simple two-node links between a sender (Alice) and a receiver (Bob), recent advances now extend this to multi-node entanglement-based quantum network architectures \cite{Hasan2023, Pompili2021, Huang2025, Lu2021, Fan2025}.

Efforts to improve compactness and modularity have also led to significant progress in using quantum technologies outside laboratories \cite{Anwar2022, Kouadou2022, Singh2025}, such as satellite-based quantum communication systems \cite{Chen2021} and mobile quantum communication links \cite{Conrad2023, Wang2026a} based on spontaneous parametric down-conversion sources of entangled photon pairs. Although such sources are relatively straightforward to implement using non-linear optical techniques, their performance is ultimately constrained by their inherent Poissonian photon-number statistics \cite{Zhang2021}. In order to realize deterministic quantum communication schemes, quantum light sources are therefore indespensible. Over the past decade, great progress towards has been made on the field of efficient and compact single photon sources using quantum emitters \cite{Somaschi2016, Snijders2018, Musial2020, Tomm2021, Northeast2021, Rickert2025}. By integrating these sources into testbeds and optical distribution systems the vision of large-scale quantum communication and specifically quantum key distribution networks has been brought closer to reality \cite{Gao2022, Liu2022, Zhang2023, Yang2024, Wang2026}.

To build multipartite quantum information-exchange networks and quantum-repeater architectures for future quantum networks byond point-to-point connections, it is necessary to use on-demand entangled multi-photon (i.e. pair) sources that are highly efficient, feature close to unity entanglement fidelities and high indistinguishabilities \cite{Kim2008, Acin2018, Loock2020,Schimpf2021,Zhang2023}. Notably, the exciton–biexciton cascade in GaAs quantum dots (QDs) provides a highly efficient mechanism for entangled-photon pair emission, positioning these systems as leading candidates for on-chip entangled photon sources \cite{Bounouar2015,Chen2018,Wang2019,Liu2019,Hopfmann2021, Li2023,Strobel2024,Zena2025,Langer2025a}.

Despite substantial progress in source physics and optical performance, the majority of demonstrated entangled photon sources remain confined to laboratory environments and are not readily deployable in applied or industrial contexts. This gap between laboratory demonstrations and field-ready systems is not primarily limited by entanglement quality or single-photon purity, but rather by system-level challenges. These include the need for continuous manual alignment, limited operational stability, bespoke experimental setups and layouts, and tight coupling between optical performance and strict environmental controls such as temperature and vibrational stability. As a result, state-of-the-art entangled photon sources are difficult to operate reliably outside controlled laboratory settings, hindering their adoption in emerging quantum communication network demonstrators that demand long-term, autonomous, and reproducible operation in industrial environments such as server rooms.

In parallel, applied quantum photonics has begun to place increasing emphasis on deployment-oriented metrics, such as operational uptime, automation, footprint, and system interoperability \cite{Eisaman2011, Sibson2017}. For entangled photon sources to transition from experimental platforms to enabling infrastructure, their design must explicitly account for these requirements at the architectural level, rather than treating them as secondary engineering considerations.

Towards the goal of industrial deployment of high-performance quantum systems based on \GaAs \, QDs, significant progress has been made by developing fiber-based platforms. Fiber-coupled source systems integrated into cryogenic environments below \qty{10}{\kelvin} are necessary to attain high-performance devices on small footprints. Such systems have demonstrated reliable and scalable operation, providing a pathway toward practical and commercial deployment. Notable advancements include the development of efficient single photon sources \cite{Tomm2021, Vajner2022}, indistinguishable single photons \cite{Huber2017},  fiber-pigtailed, integrated sources \cite{Margaria2025}, GHz-clocked entangled photon pair sources based on QDs \cite{Shooter2020, Hopfmann2021}, and ultra-compact, fiber-coupled single and entangled photon pair sources using 3D-printed micro-objectives \cite{Bremer2020, Langer2025}. Additionally, techniques for collecting telecom-wavelength photons from circular Bragg gratings with optical fibers and 3D-printed microlenses have been demonstrated \cite{Tran2025}. Together, these approaches represent a robust foundation for high-performance, fiber-integrated quantum photonic devices for semiconductor QDs suitable for industrial applications.
In this work, we address this challenge by developing and experimentally validating a rack-based, mobile entangled photon source built around a semiconductor \GaAs \, QD emitter. Rather than optimizing a single performance metric in isolation, our approach focuses on system-level integration and operational robustness, while retaining near-ideal entanglement quality. Recent studies demonstrate that relevant performance indicators for industrial quantum networks are the temporal stability of three key parameters: the photon-pair generation rate, the indistinguishability of the emitted photons, and the entanglement fidelity \cite{Margaria2025, Craddock2024}. To frame this approach, we operationally define \emph{industrial compatibility} of an entangled photon source through the following system-level properties:

\begin{itemize}
    \item \textbf{Standardized footprint:} integration within 19-inch rack enclosures, compatible with existing industrial infrastructures such as server rooms and fiber network nodes.
    \item \textbf{Automated operation:} closed-loop stabilization of optical alignment, excitation conditions, and polarization settings, with a target autonomous operation time exceeding \qty{24}{hours} without manual intervention.
    \item \textbf{Operational stability:} sustained entangled photon generation with less than \qty{5}{\percent} drift in key performance metrics (brightness, fidelity) over the autonomous operation time period.
    \item \textbf{Remote accessibility:} full software-based control and monitoring via standardized communication interfaces, enabling unattended and geographically distributed operation.
    \item \textbf{Deployability:} vibration and thermal robustness compatible with relocation without permanent degradation of source performance.
\end{itemize}

These criteria translate the often qualitative notion of deployability into concrete, testable system requirements and form the design rationale for the architecture presented in this manuscript.

Using this framework, we demonstrate polarization-entangled photon pair generation with an entanglement negativity 2n of up to \num{0.98 \pm 0.01}. Moreover, we show sustained and hands-off operation over a six-hour time window with an average single-photon emission rate of $\qty{697 \pm 8}{\kHz}$ while maintaining a 2n-value exceeding $\qty{95}{\percent}$. These results are obtained in a fully integrated, rack-based system incorporating optical excitation and collection, cryogenic operation, and automated data collection and analysis infrastructure.

The remainder of this work is structured as follows.  \cref{sec:Intro} introduces the system requirements and design rationale derived from the industrial compatibility criteria outlined above,  \cref{sec:design} discusses the system design and illustrates the architecture and implementation of the rack-based quantum dot entangled photon pair source system. In \cref{sec:source}, the entangled photon pair source with in situ fiber coupling of the quantum dot inside a compact cryostat is discussed. This is followed by \cref{sec:Result} in which experimental results validating entanglement quality, brightness, and long-term stability are presented. Finally, \cref{sec:conclusions} concludes the manuscript. Supplementary material providing additional experimental details and analysis is also included for experimental apparatus \cref{sec:experimental_apparatus}, exciton fine-structure splitting \cref{sec:exciton _Fine Structure Splitting}, correlation tomography \cref{sec:Correlation_tomography}, and lifetime \cref{sec:Lifetime}.

\section{System Design and Implementation}
\label{sec:design}

\begin{figure}[H]
    \centering
    \includegraphics[width=\textwidth]{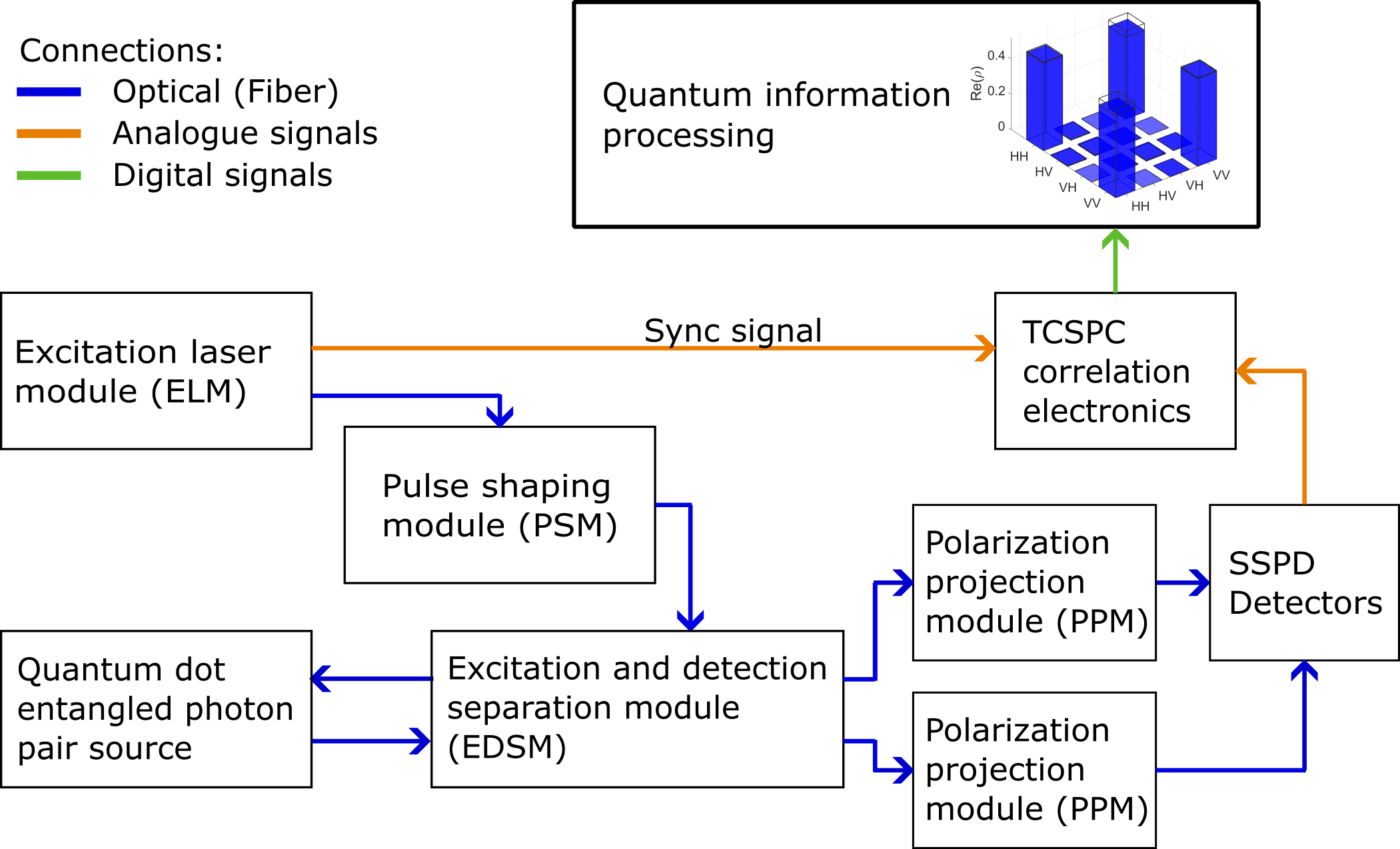}
    \caption{\label{fig:System_Overview}
High-level schematic overview of the compact, industry-compatible entangled photon pair source system architecture housed within 19-inch rack systems. The principles and design considerations of the various modules are discussed in the text.
 }
\end{figure}

The entangled photon pair source presented system in this work incorporates of all the required hardware to operate the source. As illustrated in \cref{fig:System_Overview}, this includes the cryogenically cooled \GaAs  \, QD emitter, the excitation laser system, all optics, polarization projection, the superconducting single-photon detector (SSPD) system, the time-correlated single-photon counting (TCSPC) electronics, the computer-based quantum information processing as well as the automated control system. The basic working principle of this entangled photon pair source follows that of previous successful laboratory-based demonstrations of such systems \cite{Hopfmann2021, Langer2025}.

\begin{figure}[H]
    \centering
    \includegraphics[width=\textwidth]{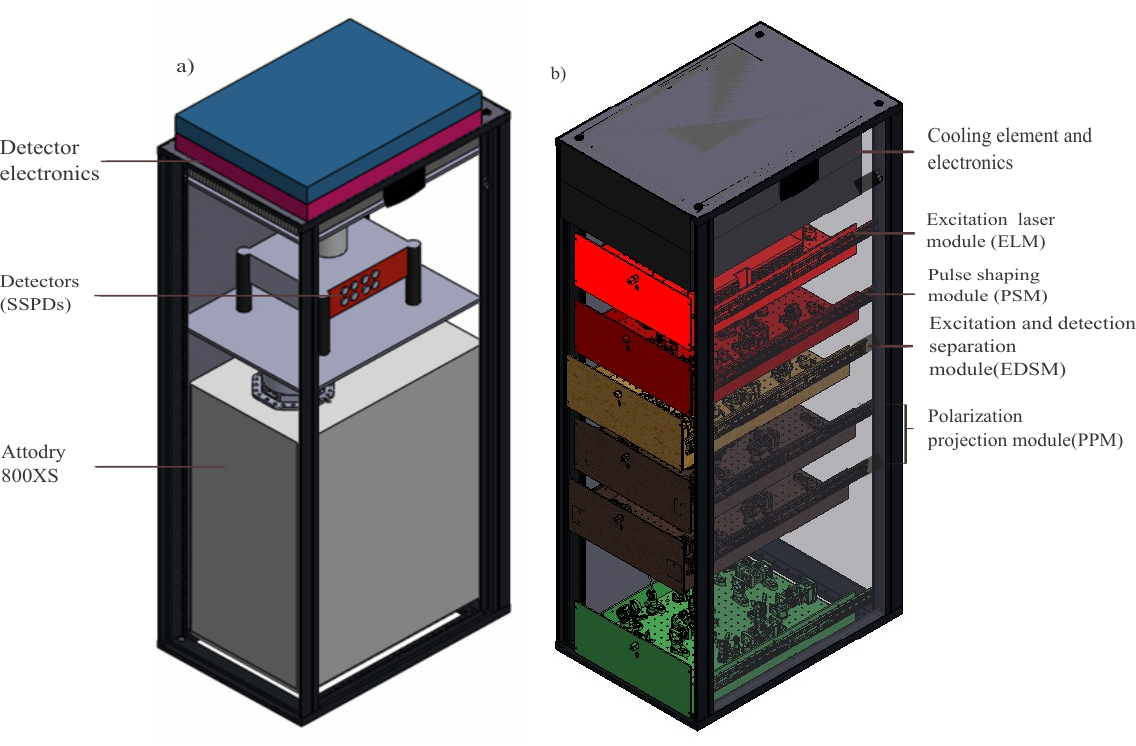}
    \caption{\label{fig:Rack_Schematic}
Illustration of the rack-based quantum dot entangled photon pair source system architecture.  (a) Source cryostat and detection rack: Contains the cryostat system together with the associated electronics for the GaAs quantum dot source and single photon time-resolved detection. (b) Optics rack: The optical sub-systems are organized into standardized modules. The automatized polarization projection units and the pulsed 1 GHz clocked Ti:Sa laser system, including its controller, are fully integrated into this rack. and see \cref{fig:System_Overview} for their interactions.
 }
\end{figure}

The source system presented here is, however, implemented within two industry-standard \qty{600}{\mm} wide, mobile 19-inch racks on wheels, as schematically illustrated in \cref{fig:Rack_Schematic}. This modular architecture allows compact integration of all components while maintaining flexibility for system modifications. Interconnections within each rack and between the two racks are realized via electrical cables and low-loss single mode optical fibers, ensuring reliable signal routing and minimal transmission losses. Due to these premises, the free-space optical excitation and detection of the QD-source inside the cryostat and therefore the employment of direct in situ fiber coupling of the QD output signal, as established in Ref. \cite{Langer2025}, becomes indispensable.\\

One rack is primarily dedicated to housing the QD-chip as well as the SSPDs each within closed-cycle cryostats. The fiber-coupled QD-chip constituting the entangled photon source is placed within a Attocube 800XS cryostat from Attocube Systems GmbH. The employed SSPDs system is commercial system with a nominal timing resolution of \qty{20}{\ps} (root-mean-square) and \qty{85}{\percent} efficiency from Single Quantum. The two rack-mountable He-compressors required for the cryogenic operation (IGLU from Attocube Systems GmbH) are placed next to the cryogenic systems rack. 


The second, designated as the optics rack, contains all optical instrumentation required for the operation of the entangled-photon pair source. This includes the excitation laser source, wavelength filtering optics, polarization projection units, and vibration dampening system. A detailed description of the employed systems and devices are found in the supplementary materials \cref{sec:experimental_apparatus}.

All optical systems are subdivided into compact modules which consist of \qty{400}{\mm} $\times$ \qty{600}{\mm} aluminium breadboards. These modules are placed on rubber vibration isolation pads in drawers of \qty{4}{U}\footnote{A standard rack height unit (U) corresponds to height of \qty{44.45}{\mm}.} height mounted within the rack. This approach facilitates both access for fine-tuning of the optics while retaining a compact footprint while in operation. For optical in- and outputs of the modules low-loss E2000 fiber connectors in conjunction with \qty{780}{\nm} single mode active core-aligned fibers (from Diamond SA) are used, which features losses \qty{<0.2}{dB} per connector. This approach ensures minimal losses of the optical module interconnects. The following optical modules are employed in this work:

\begin{itemize}
    \item Excitation laser module (ELM),
    \item Pulse shaping module (PSM),
    \item Excitation and detection separation module (EDSM),
    \item Two-polarization projection modules (PPM).    
\end{itemize}

The optical modules are mounted in a thermally isolated fully enclosed rack system. In future iterations of this system it is planned to stabilize the temperature within the rack to improve the long-term stability of the system for variable environments. This active stabilization system, however, is not used in this work. Indirect temperature stability is provided by the laboratory climate control system, which nominally provides stabilization to \qty{18 \pm 2}{\degreeCelsius}.

The QD optical excitation is provided by a GHz-clocked, \qty{40}{\ps} pulsed width Ti:Sa laser system (Novanta Taccor Tune). This laser system output features a spectral bandwidth of \qty{14}{\nm} which is filtered within the ELM using a transmission grating (\qty{1200}{l/mm}, \qty{800}{\nm} blaze) down to a spectral full width half maximum (FWHM) of \qty{0.35}{\nm} before coupling it to a polarization maintaining single mode fiber. In order to further reduce the spectral bandwidth and improve the signal to background ratio, the excitation signal is filtered in the PSM using another transmission grating to a FWHM of \qty{0.1}{\nm}. The intensity of the excitation signal can be varied continuously via a computer-controlled \qty{780}{nm} half-wave plate before a linear polarizer. The filtered excitation signal is fed into the EDSM, in which it is overlaid with the output signal of the QD and send to the QD using a volume notch filter of \qty{0.4}{\nm} bandwidth tuned to the excitation wavelength as a spectrally selective mirror. Entangled photon pairs are created from the \GaAs \, QD using the established process of two-photon excitation of the QD biexciton (XX) state \cite{Michler2003}, see \cref{sec:Result} for further details. The resulting polarization entangled QD exciton (X) and biexciton (XX) photon signals are separated and filtered in the EDSM into respective output signals. These X and XX output signals constitute the core of the entangled photon pair source and can, in future uses of the presented system, be employed as a core building block of quantum communication networks. In presented implementation, the X and XX output signals are directed to two separate PPMs. In these modules, the X and XX photons are projected using computer-controlled combination of half- and quarter-zero-order waveplates before a linear polarizer. This system enables the projection onto the linear (H/V), diagonal (D/A) and circular (R/L) polarization directions, cf. \cref{sec:Result} and our previous publications \cite{Hopfmann2021, Langer2025}. The projected photons are detected using the rack-based SSPD system, the output of which are collected as time- and channel-resolved single photon detection events using commercial time tagger electronics (TimeTagger Ultra from Swabian Instruments GmbH). Using hardware-accelerated software-based processing, the event data is turned into time-correlated single photon counting (TCSPC) signals, which are processed further in order to extract the time- and polarization-resolved two-photon correlation required to determine the two-photon density matrix of the source in a given time frame.


\section{Entangled Photon Pair Source}
\label{sec:source}
\begin{figure}[H]
    \centering
    \includegraphics[width=\textwidth]{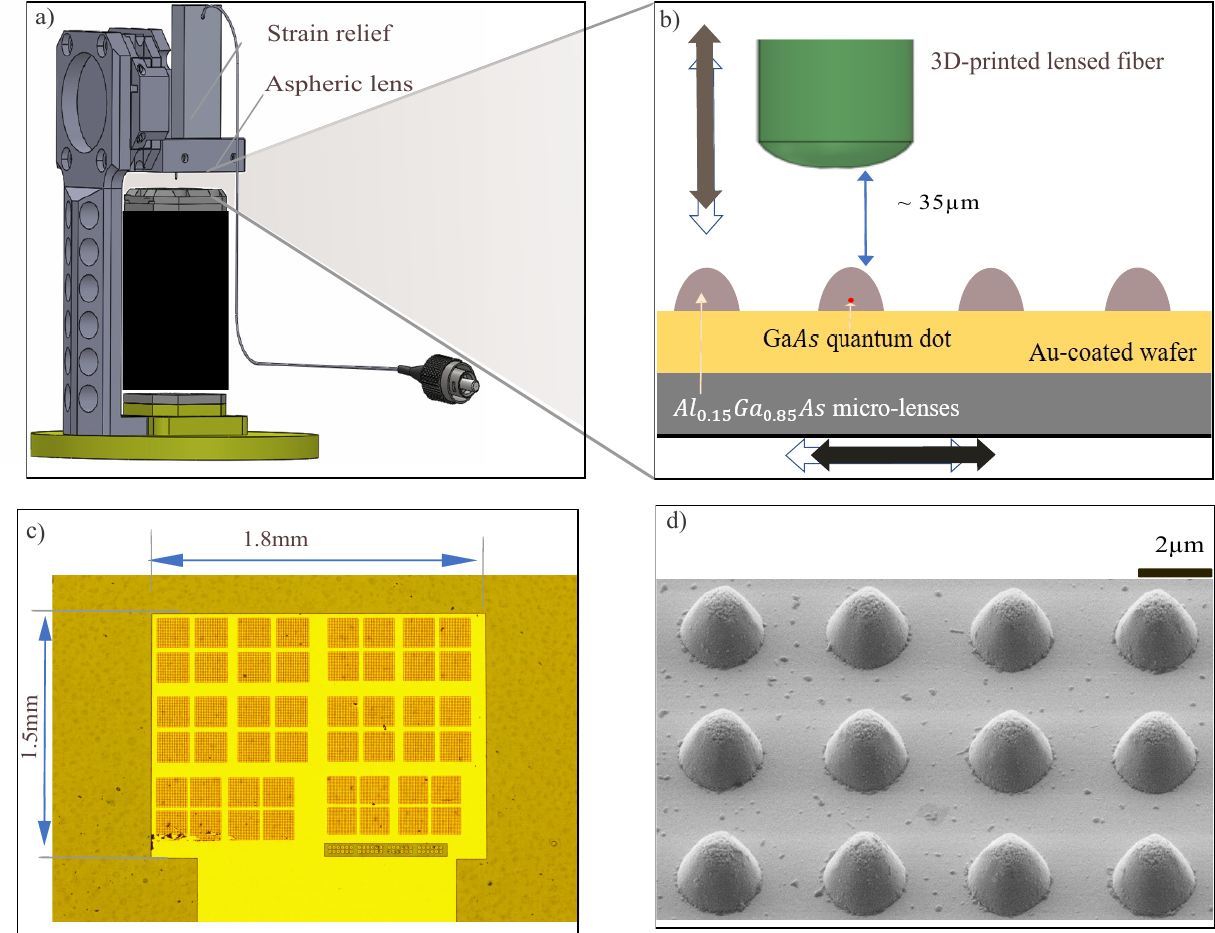}
    \caption{\label{fig:Fiber_coupling}
Overview of the employed approach of in situ fiber-coupling of entangled photon pairs sources based on GaAs quantum dots embedded in \AlGaAs \, monolithic microlenses. a) Three-dimensional representation of the cryostat interior design of the in situ fiber coupling system, which includes the fiber-strain-relief unit and free-space-accessible lens for coarse fiber–micro-objective alignment. b) Illustration of the QD-microlens emission collection into a 3D-printed micro objective attached to a single mode fiber.  c) A microscopy picture of the monolithic QD-microlens sample chip used in this study. Its dimensions is 1.8 × 1.5 mm, featuring 19200 lenses arranged in 48 fields of 20 $\times$ 20 lenses. d) Scanning electron micrograph under a \qty{45}{\degree} tilt.  }
\end{figure}

One of the key components of the entangled photon pair source is the optically driven and coupled GaAs QD. Due working principle of this type quantum emitter, cryogenic cooling to temperatures below \qty{10}{\kelvin} is indispensable \cite{Michler2003,Keil2017}. At the same time, due to the architectural requirements outlined above, in situ coupling to single-mode fibers is necessary to build a deployable system outside laboratory environments. Earlier implementations of in situ fiber coupling relied on fixed-glue assemblies \cite{Schlehahn2018, Sartison2021, Bremer2022, Rickert2025}, which provide no free mechanical alignment once cured, and only one QD can be used because of its fixed architecture. While simple, this approach is highly sensitive to thermal strain, especially because device preparation is done at room temperature while measurements are carried out in cryogenic environments. As a result, such fiber coupling approaches often exhibit limited brightness, are not reconfigurable, and typically allow only a limited number of cooling cycles.
In the present work, we extend our previous work on compact entangled photon pair sources using monolithic QD-microlenses in situ coupled to single-mode fibers with 3D-printed micro objectives \cite{Langer2025a,Langer2025} to a more compact cryostat geometry compatible with 19-inch rack systems. This approach avoids the limitations of glue-fixed assemblies and enables reproducible alignment under cryogenic conditions.




To protect the optical fiber from mechanical stress and potential damage, a strain-relief mechanism was incorporated into the system. The QD-chip was mounted on a XY translation stage beneath the fiber, enabling lateral scanning and allowing individual microlenses to be addressed as needed. Additionally, a Z-axis stage is used to control the distance between the fiber and the QD-chip, see illustration in Ref. \cref{fig:Fiber_coupling}(a). The optimal focus distance between fiber micro objective and QD-chip is about \qty{35}{\um}, illustrating the compact nature of the arrangement, cf. \cref{fig:Fiber_coupling}(b). \cref{fig:Fiber_coupling}(c) shows an optical microscopy image of the monolithic QD-microlens sample utilized in this work, the fabrication process of these devices is detailed in \cite{Langer2025a}. \cref{fig:Fiber_coupling}(d) presents a scanning electron microscopy (SEM) image, recorded at a \qty{45}{\degree} tilt, depicting an array of \AlGaAs \, monolithic microlenses QD-microlenses fabricated on a gold-coated GaAs substrate.

This compact design is integrated into the cryogenic system, as shown in \cref{fig:Rack_Schematic}(a), and is connected to the optical rack via a single mode fiber. This arrangement mechanically decouples the optical system from the cryogenic environment, significantly improving its robustness - especially outside well-controlled laboratory environments. Consequently, both the footprint and complexity of the optical and optoelectronic subsystems are significantly reduced, optimizing the use of available space and simplifying the overall setup. The optical connection is implemented using a single-mode fiber terminated with low-loss connectors, providing reliable, efficient coupling of the quantum light source to the optical system.

\section{Results and Discussion}
\label{sec:Result}
\begin{figure}[H]
    \centering
    \includegraphics[width=\textwidth]{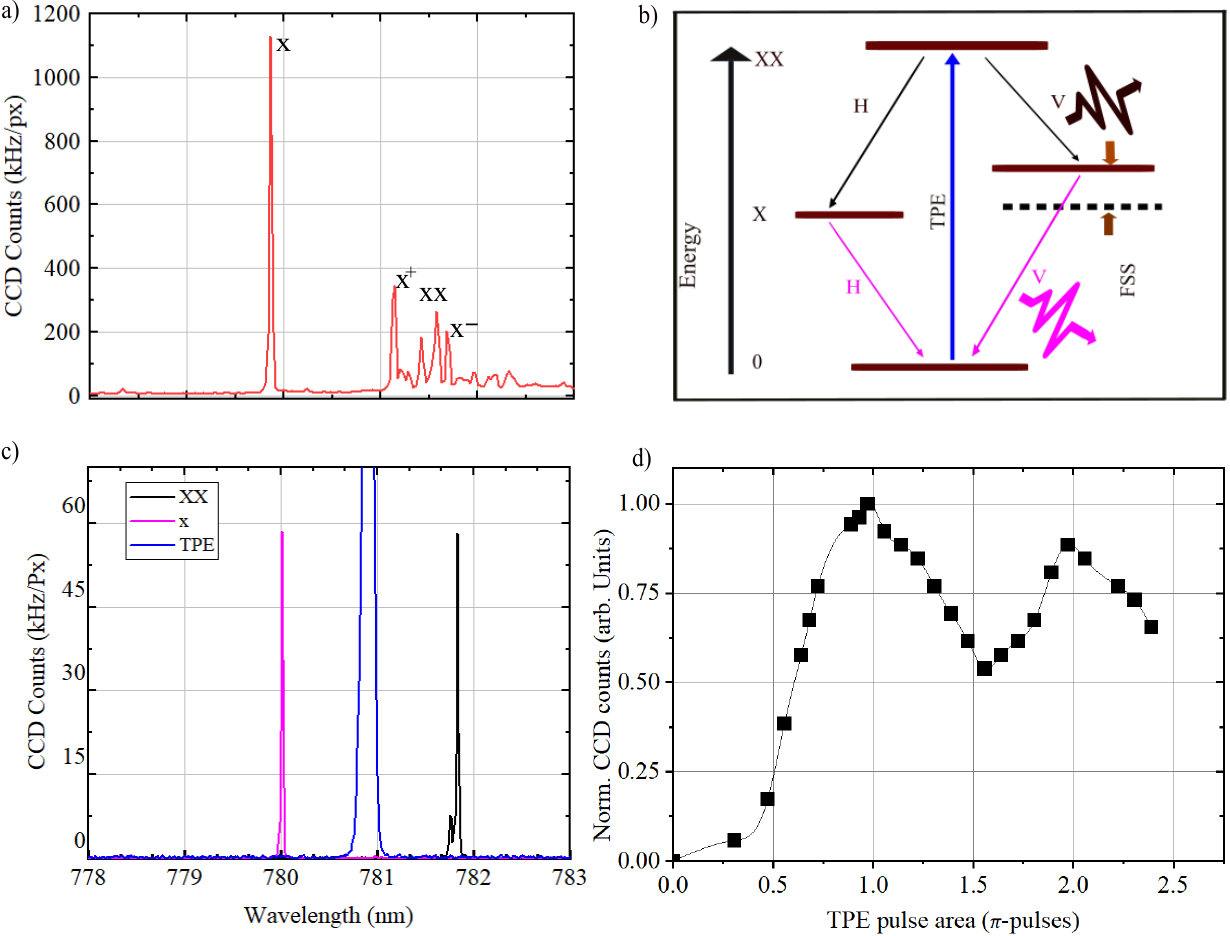}
    \caption{\label{fig:Spectroscopy}
 a) Above-band excitation spectrum of a single GaAs QD-microlens captured using a 3D-printed objective on top of a single-mode fiber. The exciton (X), biexciton (XX), negative and positive trion ($X^-$ and $X^+$) QD emission lines are annotated. b) Schematic illustration of the two-photon resonant excitation (TPE) and entangled photon pair creation scheme. c) Micro-photoluminescence spectrum of the selected QD under two-photon excitation at the $\pi$-pulse excitation. The $\lvert XX\rangle \to \lvert X\rangle$ and the $\lvert X\rangle \to \lvert 0\rangle$ emission lines are shown in back and pink, respectively. d) Normalized CCD count rate as a function of TPE pulse area. The first Rabi $\pi$-pulse is achieved at a power of \qty{9}{\micro \W} at a laser repetition rate of \qty{1}{\GHz}.}
\end{figure}

To assess the practical usability of the optical system developed in section \cref{sec:design}, it is essential to evaluate its optical performance under realistic operating conditions. For this purpose, a series of experimental measurements are carried out. In order to find  QDs in microlenses suitable as entangled photon pair sources on the sample chip, above-band photoluminescence spectroscopy is performed. The employed apparatus is presented in the supplemental materials \cref{sec:experimental_apparatus}. Criteria for the suitability of QDs for the purpose of entangled photon pair sources are mainly high brightness, exciton emission wavelength within \qtyrange{778}{781}{\nm} and low fine-structure splitting (FSS). After investigation of about 16 QD-microlenses, in this wavelength range the best QD candidate was identified, cf. \cref{fig:Spectroscopy}(a). All subsequent experimental results shown in this work are on this specific QD.
To utilize the QD as a deterministic entangled photon pair source, pulsed resonant two-photon excitation (TPE) is employed \cite{Stufler2006, Bounouar2015}. This excitation approach enables coherent resonant pumping of the XX state through the simultaneous absorption of two photons, which consequently undergoes a cascaded radiative decay that produces two photons, one from the XX $\to$ X transition and one from the X $\to$ ground-state $\lvert 0\rangle$ transition \cite{Stufler2006, Bounouar2015}, cf. \cref{fig:Spectroscopy}(b). Because angular momentum is conserved \footnote{Specifically, both the initial XX state $\lvert XX\rangle$ and the final ground state $\lvert 0\rangle$ have a total angular momentum $j$ of zero, while the emitted photons carry angular momentum $j_\nu = \pm 1$. It therefore follows that the spins (polarizations) of X and XX photons are quantum-correlated because $j^{XX}_\nu + j^X_\nu= 0$.} throughout these two sequential spontaneous emission events, the emitted photon pairs are polarization-entangled. Owing to the coherent nature of the TPE excitation process, the population of the XX state exhibits Rabi oscillations that depend on the excitation pulse power, cf. \cref{fig:Spectroscopy}(d). It can therefore be used to realize deterministic QD-entangled photon pair sources by using the first Rabi $\pi$-pulse achieved at a continuous wave excitation power of \qty{9}{\micro \W}. This corresponds to a per-pulse energy of \qty{9}{\atto \joule}.
Due to the finite FSS of the X state polarization state procession, oscillations are induced; this effect is discussed in detail in Refs. \cite{Winik2017,Hopfmann2021}. For the quantum dot studied here, the fine-structure splitting $\Delta FSS$ is determined using polarization resolved PL-spectroscopy to \qty{2.54(12)}{\micro \eV}. Details on the experimental derivation of this value is found in the supplemental material \cref{sec:exciton _Fine Structure Splitting}.



To reconstruct the two-photon density matrix of the entangled state, a full polarization tomography by measuring time-resolved X–XX coincidences across all \num{6} $\times$ \num{6} = \num{36} polarisation basis combinations is performed. Using an integration time of \qty{60}{\second \per combination} a complete tomography is performed in \qty{24}{\minute}. The presented coincidence measurements, the full matrix representation of the coincidences can be found in supplemental materials \cref{sec:Correlation_tomography}, is performed by integrating five tomography iterations representing an integration time of \qty{120}{\minute} in order to assure high statistical significance of the dataset. The two-photon density matrix reconstruction procedure follows the standard methodology described in Refs. \cite{James2001,Winik2017, Hopfmann2021}. 

\begin{figure}[H]
    \centering
    \includegraphics[width=0.8\textwidth]{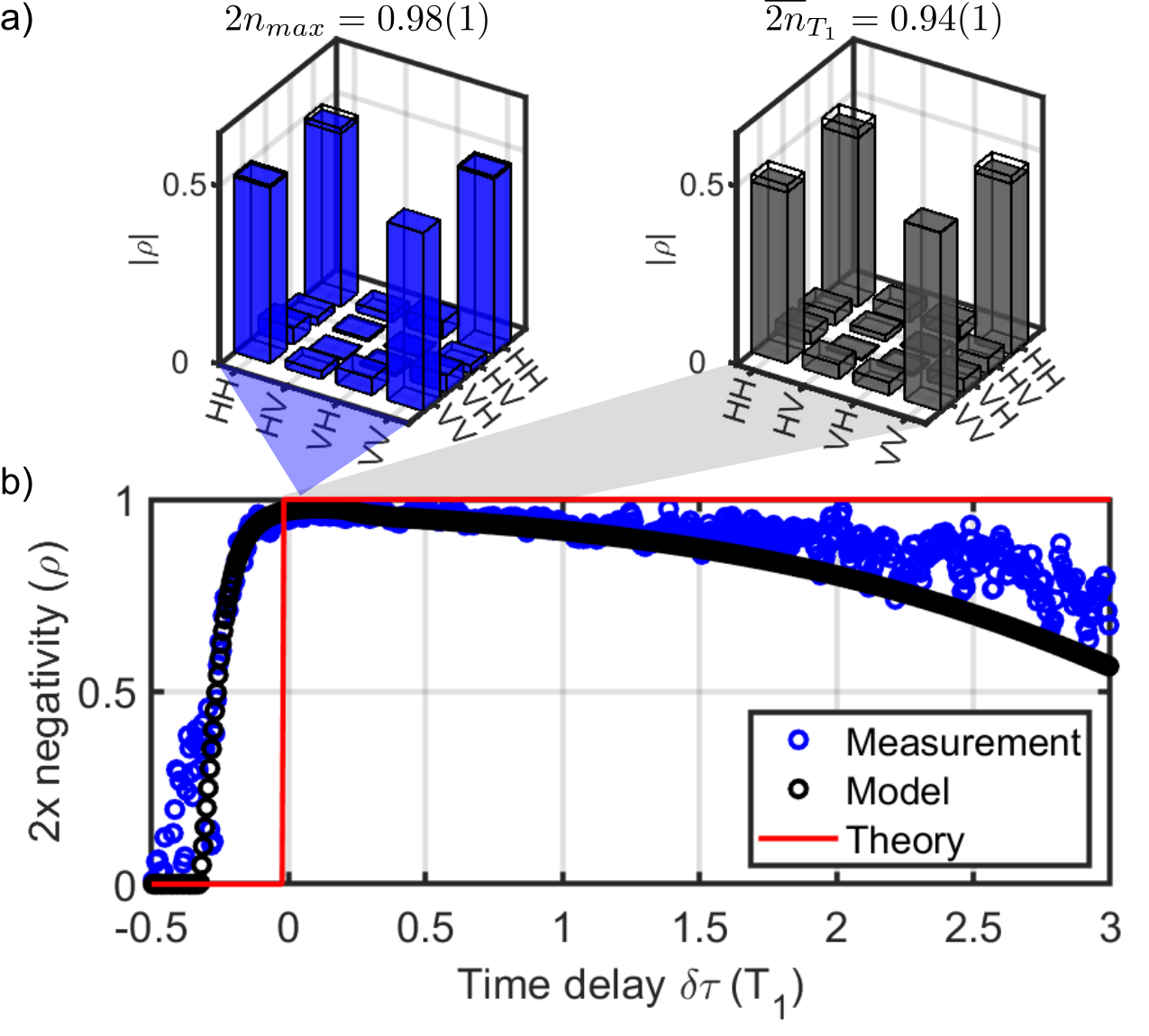}
    \caption{\label{fig:Entanglement}
        (a) Representation of the absolute values of two-photon density matrix $\rho$ using \qty{1}{\GHz}-clocked two-photon resonant excitation. Both $\rho$ representations for the maximal within a \qty{8}{\ps} window and lifetime $T_1$-averaged entanglement negativities $2n_{max}$ and $\overline{2n}_{T_1}$, respectively, are shown. (b) 2n as a function of time delay $\delta \tau$ between the X-XX two-photon coincidences. The measured $\rho$ and 2n (blue curve) are derived from the full X-XX two-photon coincidence tomography measurement over all \num{36} detection polarization combinations; the full dataset is shown explicitly in the supplementary materials \cref{fig:Two-photon X–XX coincidence}. The green and red curves represent the calculated 2n values of an ideal maximally entangled model with and without considering a limited single-photon detector timing resolution, respectively.}
\end{figure}

The time-resolved two-photon density matrix results are presented in \cref{fig:Entanglement}. Besides the experimental data shown in blue, the ideal maximally entangled theoretical model and the theoretical model with limited detector timing resolution is shown in red and black, respectively. The assumed full-width half-maximum two-photon detector timing resolution is \qty{50}{\ps}. The good agreement between the experimental data and the maximally entangled state model indicates that the entangled photon pairs generated by this source closely approximate a maximally entangled state. The top panel of \cref{fig:Entanglement} exemplarity shows the two-photon density matrix, $\mathrm{Re}(\rho)$ exhibiting the maximal entanglement negativity (2n) of \num{0.98(1)} within a \qty{8}{\ps} window. The degree of entanglement was quantified using the entanglement negativity, \qty{2}{n} as a function of the time delay $\delta \tau$ between the XX and X photon emission events in units of the X lifetime $T^X_1$. The X lifetime is determined to a value of \qty{162(4)}{\ps}. The weighted average $2n$-value over one $T^X_1$ equates to \num{0.94(1)}, demonstrating the highly entangled nature of the emitted photon pairs in the presented source even in the presence of finite FSS. Using the XX auto-correlation coincidences from the same two-photon coincidence measurement a single photon purity of \qty{99.2(2)}{\percent} of the source can be inferred.


\begin{figure}[H]
    \centering
    \includegraphics[width=\textwidth]{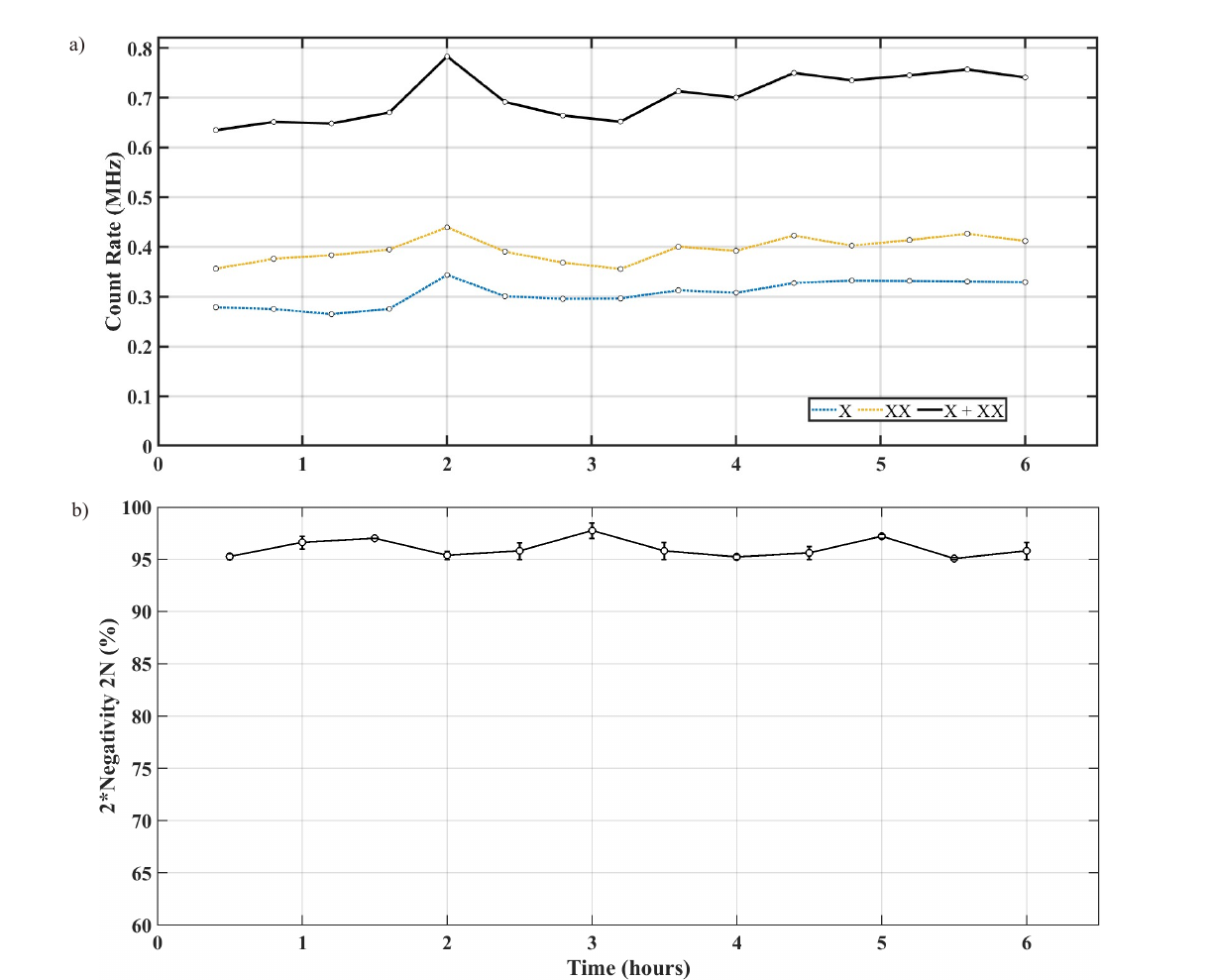}
    \caption{\label{fig:time_performance}
   Unattended performance of the rack-based entangled photon pair source over a six-hour window. a) X, XX and combined single photon detection rates collected in the V polarization basis over time. b) Entanglement negativity 2n as a function of time, error bars are indicated. 
}
\end{figure}

One of the important aspects for industry-compatible entangled photon pair sources is reliable long-term performance. In order to assess this aspect in the presented source system a six-hour uninterrupted and hands-off operation is performed. Importantly, no active polarization or temperature stabilization system is used during this measurement, this enables evaluation of the system robustness in a realistic environment.

\cref{fig:time_performance}(a) and (b) present the corresponding time traces of the single-photon count rate and the entanglement negativity. In total, the presented uninterrupted measurement incorporates \num{24} tomography iterations recorded over the six hours; two consequent iterations are used to derive the entanglement negativity over time. The average rerecorded combined X and XX single photon flux as observed in the V polarization direction is {\qty{697\pm 8}{\kHz}, the flux exhibits fluctuations of less than \qty{15}{\percent}. The maximal observed combined single photon rate is \qty{740}{\kHz}. The average 2n is \num{0.960(4)}, all observed values, however, remain above the \num{0.95} limit. This remarkable result demonstrates the high robustness of the presented rack-mounted quantum light source.
These observations confirm that the presented ultra-compact, industrial 19-inch rack entangled photon pair source is capable of delivering high-performance results over extended periods of time without human intervention. While further long-term validation over days and weeks as well as the implementation of active stabilization mechanisms, is warranted, the presented results pave the way towards this goal and bring these sources significantly closer to practical deployment.

\section{Conclusions}
\label{sec:conclusions}

In this work, we demonstrate a modular, rack-based entangled photon pair source architecture that bridges the gap between high-performance laboratory emitters and deployable industrial quantum infrastructure. By integrating a semiconductor \GaAs  \, quantum dot emitter within a standardized 19-inch rack footprint, we address the core requirements for industrial compatibility: portability, system-level integration, and operational autonomy. 

These results validate that transitioning to a compact, fiber-coupled form factor does not necessitate a compromise in the fundamental quality of the quantum light. Through the use of monolithic\AlGaAs \, monolithic microlenses and in situ 3D-printed micro-objectives, we achieved near-ideal polarization entanglement with an entanglement negativity 2n of up to \num{0.98 \pm 0.01} under resonant two-photon excitation. This high degree of entanglement is complemented by near-perfect multi-photon suppression, evidenced by a measured single-photon purity of \qty{99.2 \pm 0.2}{\percent}. Furthermore, the system demonstrated high-rate photon emission, maintaining an average single-photon flux of \qty{697\pm 8}{\kHz} during a six-hour period of entirely unattended, hands-off operation. Crucially, the observed operational stability—characterized by 2n exceeding \qty{95}{\percent} and minimal drift in emission rates—was achieved without active temperature or polarization stabilization. This inherent mechanical and thermal robustness, facilitated by the compact and modular system design, fulfils the criteria for deployability in environments such as server rooms or network nodes. While the achieved performance characteristics do not inherently exeed previously reported values of entangled photon pair sources \cite{Wang2019, Liu2019, Hopfmann2021, Langer2025}, they are very much comparable to the state-of-the-art. Realizing this in a compact, rack-based system optimzed for compatibility to industrial environments represents a significant step towards the scalable utilization of these sources in future industrial and commercial quantum communication networks.

While the current demonstration confirms the feasibility of autonomous operation over several hours, future iterations will focus on implementing the outlined closed-loop stabilization and thermal management systems to extend the autonomous window beyond \qty{24}{hours}. By successfully housing the full quantum-optical stack—from cryogenic cooling to automated polarization projection - within a mobile, industry-aligned architecture, this work provides a scalable blueprint for the next generation of entangled photon sources required for distributed quantum communication networks.

\section*{\label{acknowledgment} Acknowledgment}
We acknowledge Yana Vaynzof and Stefan Krause for valuable discussions and suggestions. This work was funded by the German Federal Ministry of Education and Research (BMBF) projects QR.X, QR.N, QUARKS, QUIET, QD-CamNetz, and Integrated3Dprint (contracts no. 16KISQ013, 16KISQ016, 16KIS2194, 16KIS1998K, 16KISQ094, 16KISQ078, and 13N16875), German Research Foundation (DFG, grant no. 431314977/GRK2642) and DFG Reinhart Koselleck-Program (CZ 55/61-1).    

\section*{\label{das} Data Availability Statement}
All data that support the findings presented in this work are available from the corresponding author upon reasonable request.

\section*{Conflict of Interest}
All authors declare that they have no conflicts of interest. 

\clearpage

\setcounter{section}{0}
\renewcommand{\thesection}{\Roman{section}}

\section*{Supplementary Information}
\label{sec:sup}
\section{Experimental Apparatus}
\phantomsection
\label{sec:experimental_apparatus}

The experimental setup is based on the  GHz-clocked, 40-ps pulsed laser, which features a high repetition rate of 1 GHz. The laser beam follows a carefully controlled optical path to ensure precise temporal and spectral characteristics. Initially, the laser output passes through a diffraction grating and a series of wave plates, including a rotating wave plate and a polarizer, which are used to finely adjust the excitation power. This arrangement produces a narrowly defined, spectrally pure laser pulse, as schematically shown in \cref{fig:integrated_19_inch_rack_system}(c). The laser output is subsequently coupled into a single-mode fiber and directed to an E2000 Simplex fiber connector on the front panel of this module. The initial pulse exhibits a full width at half maximum (FWHM) of approximately 0.35 nm. To achieve the desired pulse characteristics, further spectral filtering is applied, and the pulse is propagated over approximately 2 meters, after the grating to improve its temporal profile, ultimately producing a pulse with an FWHM of around 0.1 nm.
 Following pulse shaping, the excitation laser is routed to the optical head, also referred to as the filtering module \cref{fig:integrated_19_inch_rack_system}(e). In this module, the laser undergoes spectral filtering to selectively separate the X and XX emission lines, as well as a 630 nm line corresponding to above-band excitation. This ensures precise excitation of the quantum dot (QD) transitions while suppressing unwanted spectral components.

Detection polarization control is implemented using two polarization projection (PP) units. Each unit consists of a set of rotating half-wave$\left(\frac{\lambda}{2}\right)$ and quarter-wave $\left(\frac{\lambda}{4}\right)$
 plates optimized for 780 nm, followed by a polarizing beam splitter \cref{fig:integrated_19_inch_rack_system}(f). The PP units are carefully calibrated and validated using a polarimeter in combination with a tunable continuous-wave SIRAH Matisse laser, adjusted to the X and XX emission wavelengths for the two respective detection paths. The polarization outputs of the PP units are rotated to align the quantum dot’s polarization eigenbasis. This alignment enables accurate two-photon correlation tomography \cref{fig:Two-photon X–XX coincidence}, allowing extraction of the two-photon density matrix and detailed characterization of the entanglement properties of the emitted photon pairs \cref{fig:Entanglement}.

\begin{figure}[H]
    \centering
    \includegraphics[width=\textwidth]{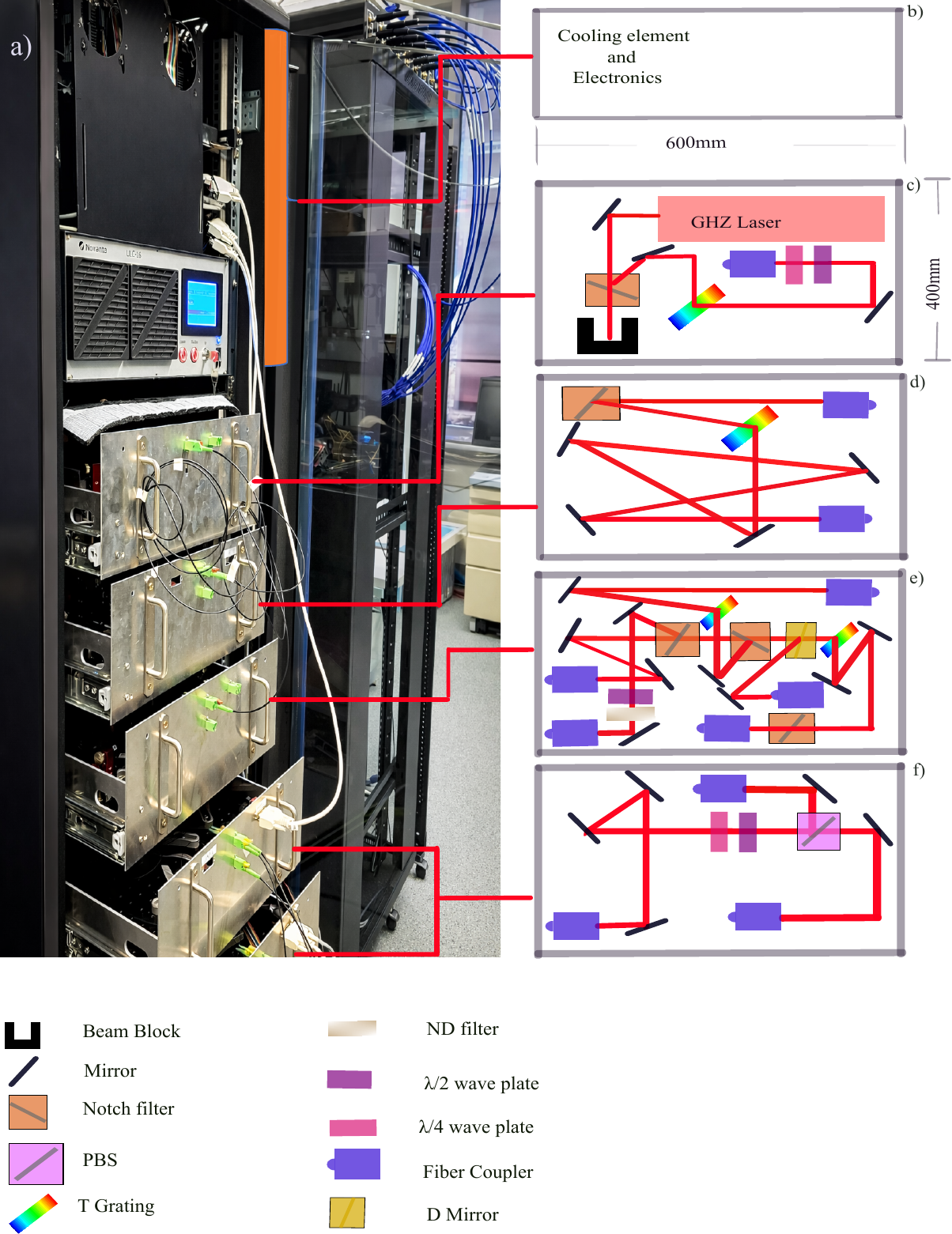}
    \caption{\label{fig:integrated_19_inch_rack_system}
a) The integrated system housed in a 19-inch rack, and
b) its associated cooling and electronics modules.
c–e) Schematic overview of the experimental setup used to characterize entanglement from GaAs quantum dots driven by resonant two-photon excitation, including the (c) Excitation laser 
module (ELM), (d) Pulse shaping Module (PSM) and (e) Excitation and detection separation module (EDSM) for separating the exciton (X), biexciton (XX) emission lines and  630 nm for above-band excitation.
 Two-photon polarization-correlation tomography (\cref{fig:Two-photon X–XX coincidence}) is performed using two or three f) polarization-projection (PP) units—depending on the number of connected nodes. Each PP unit contains rotatable $\left(\frac{\lambda}{2}\right)$
 and $\left(\frac{\lambda}{4}\right)$ waveplates. The resulting polarization states are analyzed via a polarizing beam splitter (PBS), and both output channels (H and V) are detected using superconducting single-photon detectors (SSPDs).
    }
\end{figure}

\section{Exciton Fine-Structure Splitting}
\phantomsection
\label{sec:exciton _Fine Structure Splitting}

The fine-structure splitting (FSS) of the exciton (X) manifests in the two linearly polarized eigenstates, typically denoted as
\[
|H\rangle = \frac{1}{\sqrt{2}} \left( |\uparrow\Downarrow\rangle + |\downarrow\Uparrow\rangle \right), \quad
|V\rangle = \frac{1}{\sqrt{2}} \left( |\uparrow\Downarrow\rangle - |\downarrow\Uparrow\rangle \right)
\]
where 
$|\uparrow\Downarrow\rangle, \quad |\downarrow\Uparrow\rangle$ represent the spin configurations of the electron–hole pair. Experimentally, $\Delta FSS$ can be determined by recording photoluminescence (PL) spectra as a function of the rotation angle of a half-wave plate, followed by a fixed polarizer, as illustrated in Supplementary \cref{fig: Schematic of the experimental setup}(a). The resulting spectra allow extraction of the relative emission energies of the exciton (X), biexciton (XX), and their energy difference (X–XX) as a function of polarization angle. The polarization-dependent energy shifts can be modeled to a sinusoidal function, as shown in \cref{fig: Schematic of the experimental setup}(b), yielding the fine-structure splitting of the exciton. For the quantum dot studied here, the FSS is determined to $\Delta_{\mathrm{FSS}}(X) = $ \qty{2.54 \pm 0.12}{\micro \eV}.

\begin{figure}[H]
    \centering
    \includegraphics[width=\textwidth]{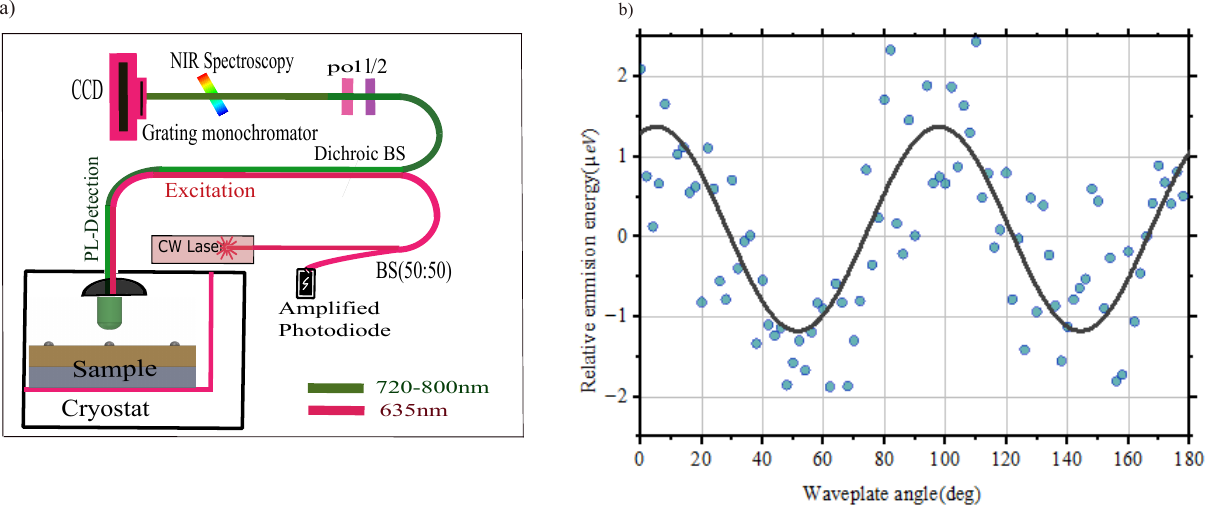}
    \caption{\label{fig: Schematic of the experimental setup}
a) Schematic of the experimental setup used for hyperspectral spatial imaging and for recording excitonic emission.
b) Exciton emission-line energies measured as a function of the $\lambda/2$
-wave plate angle in the polarization-projection unit shown in (a), a fine-structure splitting as a function of wave plate angle. 
    }
\end{figure}

\section{Correlation tomography}
\phantomsection
\label{sec:Correlation_tomography}
In the analysis of the two-photon correlations, the simulated coincidence traces are scaled solely along the coincidence axis (their y-axis) to align with the experimentally measured coincidence counts. Aside from this normalization, the model contains no adjustable or free parameters. The fine-structure splitting ($\Delta FSS$as) are independently measured, as described in section b, and are directly inserted into the theoretical model.
To reconstruct the quantum state of the photon pairs, we employ time-dependent two-photon quantum state tomography. This approach allows us to extract the full two-photon density matrix$\rho(\tau)\,\rho(\tau)\,\rho(\tau)$ as a function of the detection time delay   $\tau$ The reconstruction is performed using a maximum-likelihood estimation (MLE) procedure, following the methods established \cite{James2001,Hopfmann2021, Nie2021}.
The resulting time-resolved density matrices enable a detailed evaluation of the entanglement dynamics of the emitted photon pairs. From these matrices, we compute the entanglement negativity, which serves as a quantitative measure of bipartite entanglement. The extracted negativities, together with representative reconstructed density matrices at selected time delays, are presented in \cref{fig:Entanglement}. 

\begin{figure}[H]
    \centering
    \includegraphics[width=\textwidth]{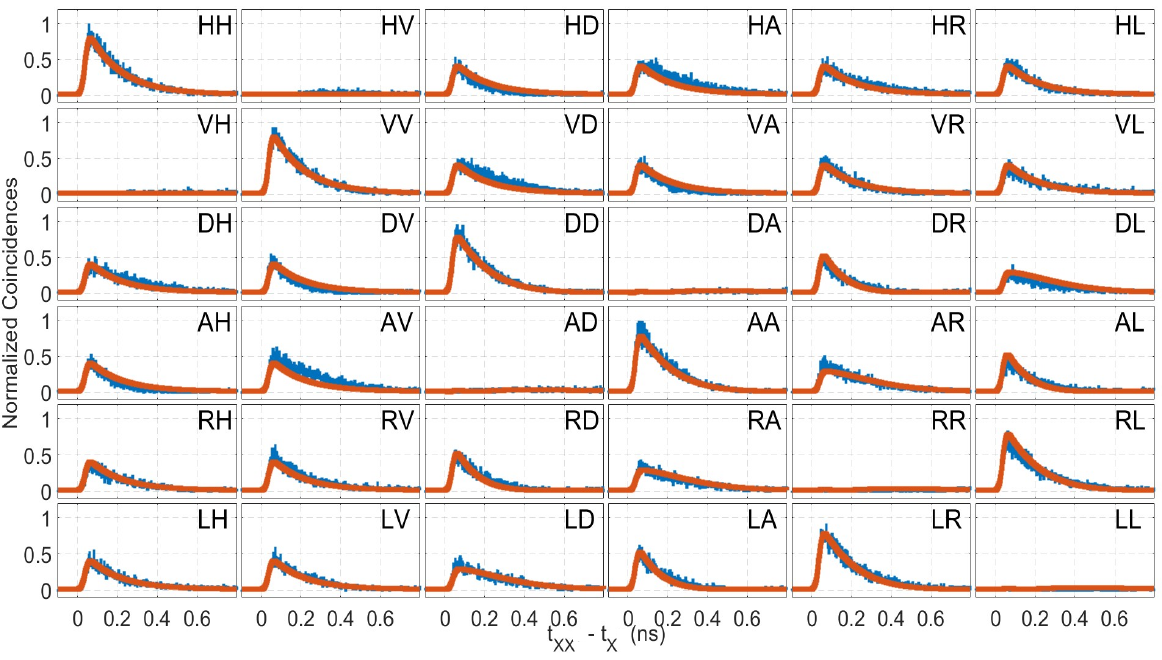}
    \caption{\label{fig:Two-photon X–XX coincidence}
Two-photon X–XX coincidence–polarization tomography as a function of the X–XX emission time delay. The blue line represents the measured data, and (dark red) line represents the model for a maximally entangled state with limited detector time resolution.}
\end{figure}
 
\section{Lifetime}
\phantomsection
\label{sec:Lifetime}

\begin{figure}[H]
    \centering
    \includegraphics[width=\textwidth]{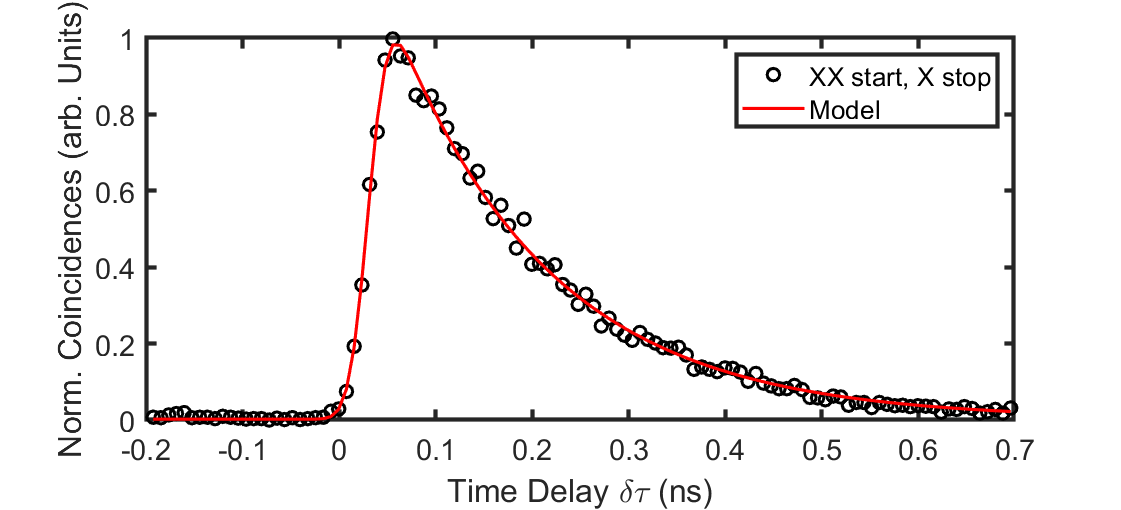}
    \caption{\label{fig:X_Lifetime}
Two-photon coincidences in the HH polarization base as a function of the time delay $\delta\tau = t_{XX} - t_X$ using a \qty{8}{\ps} time bin. Modeling by \cref{eq:X_lifetime}, see text for details, provides an X lifetime ($T^X_1$) estimate of \qty{161(4)}{\ps}.}
\end{figure}

The lifetime of QD transitions is typically inferred by recording the time-resolved photon emission correlation traces triggered by resonant excitation pulses. We employ two-photon correlation data recorded in the H polarization base using \qty{1}{\GHz} TPE to determine the X lifetime $T^X_1$. The resulting correlation trace is depicted in \cref{fig:X_Lifetime}. The correlation model functions model $C(\delta\tau)$ is given by

\begin{align}
\label{eq:X_lifetime}
	C^{X}(\delta\tau)  &\propto e^{-\delta\tau/T_1^{X}} \ast \mathcal{N}(\delta_{Det}) \, . 
\end{align}

$\mathcal{N}(\delta_{Det})$ represents a normal distribution modeling the two-photon detection time jitter of the TCSPC system. The width parameter $\delta_{Det}$ equates to a FWHM value of \qty{50}{\ps}. The two-photon event detection time delay is given by $\delta\tau = t_{XX} - t_X$. The only free parameter in the modeling process is the lifetime, which equates to $T_1^{X} = $ \qty{162 \pm 4}{\ps}.\\

\clearpage
\printbibliography

@Article{Tomm2021,
  author       = {Tomm, Natasha and Javadi, Alisa and Antoniadis, Nadia Olympia and Najer, Daniel and Löbl, Matthias Christian and Korsch, Alexander Rolf and Schott, Rüdiger and Valentin, Sascha René and Wieck, Andreas Dirk and Ludwig, Arne and Warburton, Richard John},
  date         = {2021},
  journaltitle = {Nature Nanotechnology},
  title        = {A bright and fast source of coherent single photons},
  doi          = {10.1038/s41565-020-00831-x},
  issn         = {1748-3395},
  number       = {4},
  pages        = {399--403},
  url          = {https://doi.org/10.1038/s41565-020-00831-x},
  volume       = {16},
  journal      = {Nature Nanotechnology},
  refid        = {Tomm2021},
  year         = {2021},
}

@Article{Shooter2020,
  author    = {Ginny Shooter and Zi-Heng Xiang and Jonathan R. A M\"{u}ller and Joanna Skiba-Szymanska and Jan Huwer and Jonathan Griffiths and Thomas Mitchell and Matthew Anderson and Tina M\"{u}ller and Andrey B. Krysa and R. Mark Stevenson and Jon Heffernan and David A. Ritchie and Andrew J. Shields},
  title     = {1{GHz} clocked distribution of electrically generated entangled photon pairs},
  doi       = {10.1364/OE.405466},
  number    = {24},
  pages     = {36838--36848},
  url       = {http://www.opticsexpress.org/abstract.cfm?URI=oe-28-24-36838},
  volume    = {28},
  journal   = {Opt. Express},
  month     = {11},
  publisher = {OSA},
  year      = {2020},
}

@Article{Schimpf2021,
  author       = {Schimpf, Christian and Reindl, Marcus and Basso Basset, Francesco and Jöns, Klaus D. and Trotta, Rinaldo and Rastelli, Armando},
  date         = {2021-03},
  journaltitle = {Applied Physics Letters},
  title        = {{Quantum dots as potential sources of strongly entangled photons: Perspectives and challenges for applications in quantum networks}},
  doi          = {10.1063/5.0038729},
  issn         = {0003-6951},
  number       = {10},
  pages        = {100502},
  url          = {https://doi.org/10.1063/5.0038729},
  volume       = {118},
  journal      = {Applied Physics Letters},
  year         = {2021},
}

@Article{Bounouar2015,
  author       = {Bounouar, S. and M\"uller, M. and Barth, A. M. and Gl\"assl, M. and Axt, V. M. and Michler, P.},
  date         = {2015-04},
  journaltitle = {Phys. Rev. B},
  title        = {Phonon-assisted robust and deterministic two-photon biexciton preparation in a quantum dot},
  doi          = {10.1103/PhysRevB.91.161302},
  issue        = {16},
  pages        = {161302},
  url          = {https://link.aps.org/doi/10.1103/PhysRevB.91.161302},
  volume       = {91},
  journal      = {Phys. Rev. B},
  numpages     = {5},
  publisher    = {American Physical Society},
  year         = {2015},
}

@Article{Huang2025,
  author  = {Huang, Heming and Alkhazragi, Omar and Liang, Di and Grillot, Frédéric},
  title   = {Future roles of solid-state quantum dot light sources},
  doi     = {10.1063/5.0251447},
  issn    = {0003-6951},
  number  = {8},
  pages   = {080501},
  url     = {https://doi.org/10.1063/5.0251447},
  volume  = {126},
  journal = {Applied Physics Letters},
  month   = {02},
  year    = {2025},
}

@Article{Pompili2021,
  author       = {Pompili, M. and Hermans, S. L. N. and Baier, S. and Beukers, H. K. C. and Humphreys, P. C. and Schouten, R. N. and Vermeulen, R. F. L. and Tiggelman, M. J. and dos Santos Martins, L. and Dirkse, B. and Wehner, S. and Hanson, R.},
  date         = {2021},
  journaltitle = {Science},
  title        = {Realization of a multinode quantum network of remote solid-state qubits},
  doi          = {10.1126/science.abe8798},
  issn         = {1095-9203},
  number       = {6539},
  pages        = {259--264},
  url          = {https://doi.org/10.1126/science.abe8798},
  volume       = {372},
  journal      = {Science},
  refid        = {Pompili2021},
  year         = {2021},
}

@Article{Hasan2023,
  author       = {Hasan, Syed Rakib and Chowdhury, Mostafa Zaman and Sayem, Mohammad and Jang, Yeong Min},
  date         = {2023},
  journaltitle = {IEEE Access},
  title        = {Quantum Communication Systems: Vision, Protocols, Applications, and Challenges},
  doi          = {10.1109/ACCESS.2023.3244395},
  issn         = {2169-3536},
  pages        = {1--20},
  url          = {https://doi.org/10.1109/ACCESS.2023.3244395},
  volume       = {11},
  journal      = {IEEE Access},
  refid        = {Hasan2023},
  year         = {2023},
}

@Article{Zhang2023,
  author       = {Zheshen Zhang and You, Chenglong and Maga\~na‑Loaiza, Omar S. and Fickler, Robert and de J. León‑Montiel, Roberto and Torres, Juan P. and Humble, Travis and Liu, Shuai and Xia, Yi and Zhuang, Quntao},
  date         = {2023},
  journaltitle = {Entanglement‑Based Quantum Information Technology},
  title        = {Entanglement‑Based Quantum Information Technology},
  doi          = {10.1364/AOP.497143},
  issn         = {1943-8206},
  pages        = {60--162},
  url          = {https://doi.org/10.1364/AOP.497143},
  volume       = {16},
  number       = {1},
  journal      = {Advances in Optics and Photonics},
  refid        = {Zhang2023},
  year         = {2023},
}

@Article{Acin2018,
  author       = {Acín, Antonio and Bloch, Immanuel and Buhrman, Harry and Calarco, Tommaso and Eichler, Christopher and Eisert, Jens and Esteve, Daniel and Gisin, Nicolas and Glaser, Steffen J. and Jelezko, Fedor and Kuhr, Stefan and Lewenstein, Maciej and Riedel, Max F. and Schmidt, Piet O. and Thew, Rob and Wallraff, Andreas and Walmsley, Ian and Wilhelm, Frank K.},
  date         = {2018},
  journaltitle = {New Journal of Physics},
  title        = {The quantum technologies roadmap: a European community view},
  doi          = {10.1088/1367-2630/aad1ea},
  issn         = {1367-2630},
  number       = {8},
  pages        = {080201},
  url          = {https://doi.org/10.1088/1367-2630/aad1ea},
  volume       = {20},
  journal      = {New J. Phys.},
  refid        = {Acin2018},
  year         = {2018},
}

@InProceedings{Kouadou2022,
  author       = {Kouadou, Tiphaine and Lualdi, Colin P. and Johnson, Spencer and Meier, Kristina and Aller, Josh and Slezak, Brad and Roberts, Tony and Battle, Phil and Kwiat, Paul G.},
  title        = {Compact entanglement sources for portable quantum information platforms},
  booktitle    = {Quantum Computing, Communication, and Simulation II},
  series       = {Proceedings of SPIE},
  volume       = {12015},
  pages        = {120150D},
  year         = {2022},
  doi          = {10.1117/12.2609989},
  eventtitle   = {SPIE OPTO 2022},
  eventdate    = {2022},
  publisher    = {SPIE},
  refid        = {Kouadou2022}
}

@InProceedings{Conrad2023,
  author       = {Conrad, Andrew and Isaac, Samantha and Cochran, Roderick and Sanchez-Rosales, Daniel and Rezaei, Tahereh and Javid, Timur and Schroeder, A. J. and Golba, Grzegorz and Gauthier, Daniel and Kwiat, Paul},
  title        = {Drone-based quantum communication links},
  booktitle    = {Quantum Computing, Communication, and Simulation III},
  series       = {Proceedings of SPIE},
  volume       = {12446},
  pages        = {124460H},
  year         = {2023},
  doi          = {10.1117/12.2647923},
  eventtitle   = {SPIE Quantum West 2023},
  eventdate    = {2023},
  publisher    = {SPIE},
  refid        = {Conrad2023}
}

@Article{Chen2021,
  author       = {Chen, Yu-Ao and Zhang, Qiang and Chen, Teng-Yun and Cai, Wen-Qi and Liao, Sheng-Kai and Zhang, Jin-Gang and Li, Jian and Ren, Ji-Gang and Yin, Juan and Shen, Qi and Zhou, Huadong and Liu, Xingyuan and Wang, Wei-Yue and Zhang, Sheng and Peng, Cheng-Zhi and Pan, Jian-Wei},
  title        = {An integrated space-to-ground quantum communication network over 4,600 kilometres},
  journal      = {Nature},
  volume       = {589},
  number       = {7841},
  pages        = {214--219},
  year         = {2021},
  doi          = {10.1038/s41586-020-03093-8},
  url          = {https://doi.org/10.1038/s41586-020-03093-8},
  issn         = {1476-4687},
  refid        = {Chen2021}
}

@Article{Zhang2021,
  author   = {Zhang, Chao and Huang, Yun-Feng and Liu, Bi-Heng and Li, Chuan-Feng and Guo, Guang-Can},
  title    = {Spontaneous Parametric Down-Conversion Sources for Multiphoton Experiments},
  doi      = {10.1002/qute.202000132},
  number   = {5},
  pages    = {2000132},
  url      = {https://advanced.onlinelibrary.wiley.com/doi/abs/10.1002/qute.202000132},
  volume   = {4},
  journal  = {Advanced Quantum Technologies},
  year     = {2021},
}

@Article{Kim2008,
  author    = {Kim, Danny and Economou, Sophia E. and B\ifmmode \u{a}\else \u{a}\fi{}descu, \ifmmode \mbox{\c{S}}\else \c{S}\fi{}tefan C. and Scheibner, Michael and Bracker, Allan S. and Bashkansky, Mark and Reinecke, Thomas L. and Gammon, Daniel},
  title     = {Optical Spin Initialization and Nondestructive Measurement in a Quantum Dot Molecule},
  doi       = {10.1103/PhysRevLett.101.236804},
  issue     = {23},
  pages     = {236804},
  url       = {https://link.aps.org/doi/10.1103/PhysRevLett.101.236804},
  volume    = {101},
  journal   = {Phys. Rev. Lett.},
  month     = {12},
  numpages  = {4},
  publisher = {American Physical Society},
  year      = {2008},
}

@article{Tran2025,
  author    = {Nam Tran and Pavel Ruchka and Sara Jakovljevic and Benjamin Breiholz and Peter Gierß and Ponraj Vijayan and Carlos Eduardo Jimenez and Alois Herkommer and Michael Jetter and Simone Luca Portalupi and Harald Giessen and Peter Michler},
  title     = {Collecting Telecom Photons From Circular Bragg Gratings Using Optical Fibers and 3D Printed Micro-Lenses},
  journal   = {Advanced Quantum Technologies},
  year      = {2025},
  volume    = {},
  number    = {},
  pages     = {},
  doi       = {10.1002/qute.202500450},
  url       = {https://www.advquantumtech.com/doi/10.1002/qute.202500450},
  note      = {Open access under Creative Commons Attribution License}
}

@article{Margaria2025,
  author    = {Nico Margaria and Florian Pastier and Thinhinane Bennour and Marie Billard and Edouard Ivanov and William Hease and Petr Stepanov and Albert F. Adiyatullin and Raksha Singla and Mathias Pont and Maxime Descampeaux and Alice Bernard and Anton Pishchagin and Martina Morassi and Aristide Lemaître and Thomas Volz and Valérian Giesz and Niccolo Somaschi and Nicolas Maring and Sébastien Boissier and Thi Huong Au and Pascale Senellart},
  title     = {Efficient fibre-pigtailed source of indistinguishable single photons},
  journal   = {Nature Communications},
  year      = {2025},
  volume    = {16},
  number    = {1},
  pages     = {7553},
  doi       = {10.1038/s41467-025-62712-y},
  url       = {https://doi.org/10.1038/s41467-025-62712-y}
}

@article{Huber2017,
  author    = {Daniel Huber and Marcus Reindl and Yongheng Huo and Huiying Huang and Johannes S. Wildmann and Oliver G. Schmidt and Armando Rastelli and Rinaldo Trotta},
  title     = {Highly indistinguishable and strongly entangled photons from symmetric GaAs quantum dots},
  journal   = {Nature Communications},
  year      = {2017},
  volume    = {8},
  number    = {1},
  pages     = {15506},
  doi       = {10.1038/ncomms15506},
  url       = {https://doi.org/10.1038/ncomms15506},
  note      = {Open access}
}

@article{Fan2025,
  author    = {Yun-Ru Fan and Yue Luo and Kai Guo and Jin-Peng Wu and Hong Zeng and Guang-Wei Deng and You Wang and Hai-Zhi Song and Zhen Wang and Li-Xing You and Guang-Can Guo and Qiang Zhou},
  title     = {Quantum entanglement network enabled by a state-multiplexing quantum light source},
  journal   = {Light: Science \& Applications},
  year      = {2025},
  volume    = {14},
  number    = {},
  pages     = {189},
  doi       = {10.1038/s41377-025-01805-1},
  url       = {https://doi.org/10.1038/s41377-025-01805-1},
  note      = {Open access}
}

@article{Singh2025,
  author    = {Siddhant Singh and Fenglei Gu and Sébastian de Bone and Eduardo Villaseñor and David Elkouss and Johannes Borregaard},
  title     = {Modular architectures and entanglement schemes for error-corrected distributed quantum computation},
  journal   = {npj Quantum Information},
  year      = {2025},
  volume    = {},
  number    = {},
  pages     = {},
  doi       = {10.1038/s41534-025-01146-2},
  url       = {https://doi.org/10.1038/s41534-025-01146-2}
}

@Article{Sartison2021,
  author  = {Sartison, Marc and Weber, Ksenia and Thiele, Simon and Bremer, Lucas and Fischbach, Sarah and Herzog, Thomas and Kolatschek, Sascha and Jetter, Michael and Reitzenstein, Stephan and Herkommer, Alois and Michler, Peter and Portalupi, Simone Luca and Giessen, Harald},
  title   = {3D Printed Micro-Optics for Quantum Technology: Optimised Coupling of Single Quantum Dot Emission into a Single-Mode Fibre},
  doi     = {10.37188/lam.2021.006},
  issn    = {2689-9620},
  number  = {LAM2020060010},
  pages   = {103},
  url     = {https://www.light-am.com/article/id/1ef2edd1-9aff-41d9-b155-016e71a5ae72},
  volume  = {2},
  journal = {Light: Advanced Manufacturing},
  year    = {2021},
}

@article{Schlehahn2018,
  author       = {Schlehahn, Alexander and Fischbach, Sarah and Schmidt, Ronny and Kaganskiy, Arsenty 
                  and Strittmatter, André and Rodt, Sven and Heindel, Tobias and Reitzenstein, Stephan},
  title        = {A Stand-Alone Fiber-Coupled Single-Photon Source},
  journal      = {Scientific Reports},
  volume       = {8},
  number       = {1},
  pages        = {1340},
  year         = {2018},
  issn         = {2045-2322},
  doi          = {10.1038/s41598-017-19049-4},
  url          = {https://doi.org/10.1038/s41598-017-19049-4},
}

@Article{Craddock2024,
  author    = {Craddock, Alexander N. and Lazenby, Anne and Portmann, Gabriel Bello and Sekelsky, Rourke and Flament, Mael and Namazi, Mehdi},
  title     = {Automated Distribution of Polarization-Entangled Photons Using Deployed New York City Fibers},
  doi       = {10.1103/PRXQuantum.5.030330},
  issue     = {3},
  pages     = {030330},
  url       = {https://link.aps.org/doi/10.1103/PRXQuantum.5.030330},
  volume    = {5},
  journal   = {PRX Quantum},
  month     = {8},
  numpages  = {7},
  publisher = {American Physical Society},
  year      = {2024},
}

@article{Strobel2024,
  title   = {High-fidelity distribution of triggered polarization-entangled telecom photons via a 36 km intra-city fiber network},
  author  = {Strobel, T. and Kazmaier, S. and Bauer, T. and Schäfer, M. and Choudhary, A. and Sharma, N. L. and Joos, R. and Nawrath, C. and Weber, J. H. and Nie, W. and Bhayani, G. and Wagner, L. and Bisquerra, A. and Geitz, M. and Braun, R.-P. and Hopfmann, C. and Portalupi, S. L. and Becher, C. and Michler, P.},
  journal = {Optica Quantum},
  volume  = {2},
  number  = {4},
  pages   = {274--281},
  year    = {2024},
  month   = {8},
  publisher = {Optica Publishing Group},
  doi     = {10.1364/OPTICAQ.530838},
  url     = {https://opg.optica.org/opticaq/abstract.cfm?URI=opticaq-2-4-274},
}

@article{Nie2021,
  author       = {Nie, W. and Sharma, N. L. and Weigelt, C. and Keil, R. and Yang, J. and Ding, F. and Hopfmann, C. and Schmidt, O. G.},
  title        = {Experimental Optimization of the Fiber Coupling Efficiency of GaAs Quantum Dot-Based Photon Sources},
  journal      = {Applied Physics Letters},
  volume       = {119},
  number       = {24},
  pages        = {244003},
  year         = {2021},
  month        = {12},
  issn         = {0003-6951},
  doi          = {10.1063/5.0059310},
  url          = {https://doi.org/10.1063/5.0059310},
}

@Article{Langer2025a,
  author  = {Langer, Moritz and Dhurjati, Sai A. and Zena, Yared G. and Rahimi, Ahmad and Pal, Mandira and Raith, Liesa and Nestler, Sandra and Bassoli, Riccardo and Fitzek, Frank H. P. and Schmidt, Oliver G. and Hopfmann, Caspar},
  title   = {Bright Quantum Dot Light Sources Using Monolithic Microlenses on Gold Back-Reflectors},
  doi     = {10.1088/1361-6528/add350},
  number  = {225301},
  pages   = {13},
  url     = {https://doi.org/10.1088/1361-6528/add350},
  volume  = {36},
  journal = {Nanotechnology},
  year    = {2025},
}

@inproceedings{Zena2025,
  author       = {Zena, Yared G. and Pal, Mandira and Langer, Moritz and Sai, Dhurjati and Rahimi, Ahmad and Bassoli, Riccardo and Bayelgn, Abebu A. and Czarske, Juergen and Hopfmann, Caspar},
  title        = {Enhanced emission of GaAs quantum dots in bend nanomembranes},
  booktitle    = {Quantum Communications and Quantum Imaging XXIII},
  volume       = {13618},
  pages        = {136180J},
  year         = {2025},
  doi          = {10.1117/12.3065657},
  url          = {https://doi.org/10.1117/12.3065657}
}

@article{Bremer2022,
  author       = {Bremer, Lucas and Rodt, Sven and Reitzenstein, Stephan},
  title        = {Fiber-coupled quantum light sources based on solid-state quantum emitters},
  journal      = {Materials for Quantum Technology},
  volume       = {2},
  pages        = {042002},
  year         = {2022},
  doi          = {10.1088/2633-4356/aca3f3},
  url          = {https://doi.org/10.1088/2633-4356/aca3f3}
}

@Article{Vajner2022,
  author   = {Vajner, Daniel A. and Rickert, Lucas and Gao, Timm and Kaymazlar, Koray and Heindel, Tobias},
  title    = {Quantum Communication Using Semiconductor Quantum Dots},
  doi      = {https://doi.org/10.1002/qute.202100116},
  number   = {7},
  pages    = {2100116},
  url      = {https://advanced.onlinelibrary.wiley.com/doi/abs/10.1002/qute.202100116},
  volume   = {5},
  journal  = {Advanced Quantum Technologies},
  year     = {2022},
}

@Article{Stufler2006,
  author    = {Stufler, S. and Machnikowski, P. and Ester, P. and Bichler, M. and Axt, V. M. and Kuhn, T. and Zrenner, A.},
  title     = {Two-photon Rabi oscillations in a single ${\mathrm{In}}_{x}{\mathrm{Ga}}_{1\ensuremath{-}x}\mathrm{As}/\mathrm{Ga}\mathrm{As}$ quantum dot},
  doi       = {10.1103/PhysRevB.73.125304},
  issue     = {12},
  pages     = {125304},
  url       = {https://link.aps.org/doi/10.1103/PhysRevB.73.125304},
  volume    = {73},
  journal   = {Phys. Rev. B},
  month     = {3},
  numpages  = {7},
  publisher = {American Physical Society},
  year      = {2006},
}

@Article{Somaschi2016,
  author       = {Somaschi, N. and Giesz, V. and De Santis, L. and Loredo, J. C. and Almeida, M. P. and Hornecker, G. and Portalupi, S. L. and Grange, T. and Antón, C. and Demory, J. and Gómez, C. and Sagnes, I. and Lanzillotti-Kimura, N. D. and Lemaítre, A. and Auffeves, A. and White, A. G. and Lanco, L. and Senellart, P.},
  date         = {2016},
  journaltitle = {Nature Photonics},
  title        = {Near-optimal single-photon sources in the solid state},
  doi          = {10.1038/nphoton.2016.23},
  issn         = {1749-4893},
  number       = {5},
  pages        = {340--345},
  url          = {https://doi.org/10.1038/nphoton.2016.23},
  volume       = {10},
  journal      = {Nature Photonics},
  refid        = {Somaschi2016},
  year         = {2016},
}

@Article{Chen2018,
  author  = {Chen, Yan and Zopf, Michael and Keil, Robert and Ding, Fei and Schmidt, Oliver G.},
  title   = {Highly-efficient extraction of entangled photons from quantum dots using a broadband optical antenna},
  doi     = {10.1038/s41467-018-05456-2},
  issn    = {2041-1723},
  number  = {1},
  pages   = {2994},
  url     = {https://doi.org/10.1038/s41467-018-05456-2},
  volume  = {9},
  journal = {Nat. Commun.},
  year    = {2018},
}

@Article{Wang2019,
  author    = {Wang, Hui and Hu, Hai and Chung, T.-H. and Qin, Jian and Yang, Xiaoxia and Li, J.-P. and Liu, R.-Z. and Zhong, H.-S. and He, Y.-M. and Ding, Xing and Deng, Y.-H. and Dai, Qing and Huo, Y.-H. and H\"ofling, Sven and Lu, Chao-Yang and Pan, Jian-Wei},
  title     = {On-Demand Semiconductor Source of Entangled Photons Which Simultaneously Has High Fidelity, Efficiency, and Indistinguishability},
  doi       = {10.1103/PhysRevLett.122.113602},
  issue     = {11},
  pages     = {113602},
  url       = {https://link.aps.org/doi/10.1103/PhysRevLett.122.113602},
  volume    = {122},
  journal   = {Phys. Rev. Lett.},
  month     = {3},
  numpages  = {6},
  publisher = {American Physical Society},
  year      = {2019},
}

@Article{Liu2019,
  author  = {Liu, Jin and Su, Rongbin and Wei, Yuming and Yao, Beimeng and Silva, Saimon Filipe Covre da and Yu, Ying and Iles-Smith, Jake and Srinivasan, Kartik and Rastelli, Armando and Li, Juntao and Wang, Xuehua},
  title   = {A solid-state source of strongly entangled photon pairs with high brightness and indistinguishability},
  doi     = {10.1038/s41565-019-0435-9},
  issn    = {1748-3395},
  number  = {6},
  pages   = {586--593},
  url     = {https://doi.org/10.1038/s41565-019-0435-9},
  volume  = {14},
  journal = {Nature Nanotechnology},
  refid   = {Liu2019},
  year    = {2019},
}

@Article{Lu2021,
  author       = {Lu, Chao-Yang and Pan, Jian-Wei},
  date         = {2021},
  journaltitle = {Nature Nanotechnology},
  title        = {Quantum-dot single-photon sources for the quantum internet},
  doi          = {10.1038/s41565-021-01033-9},
  issn         = {1748-3395},
  number       = {12},
  pages        = {1294--1296},
  url          = {https://doi.org/10.1038/s41565-021-01033-9},
  volume       = {16},
  journal      = {Nature Nanotechnology},
  refid        = {Lu2021},
  year         = {2021},
}

@Article{Hopfmann2021,
  author    = {Hopfmann, Caspar and Nie, Weijie and Sharma, Nand Lal and Weigelt, Carmen and Ding, Fei and Schmidt, Oliver G.},
  title     = {Maximally entangled and gigahertz-clocked on-demand photon pair source},
  doi       = {10.1103/PhysRevB.103.075413},
  issue     = {7},
  pages     = {075413},
  url       = {https://link.aps.org/doi/10.1103/PhysRevB.103.075413},
  volume    = {103},
  journal   = {Phys. Rev. B},
  month     = {2},
  numpages  = {7},
  publisher = {American Physical Society},
  year      = {2021},
}

@Article{Loock2020,
  author   = {van Loock, Peter and Alt, Wolfgang and Becher, Christoph and Benson, Oliver and Boche, Holger and Deppe, Christian and Eschner, Jürgen and Höfling, Sven and Meschede, Dieter and Michler, Peter and Schmidt, Frank and Weinfurter, Harald},
  title    = {Extending {Q}uantum {L}inks: {M}odules for {F}iber- and {M}emory-{B}ased {Q}uantum {R}epeaters},
  doi      = {https://doi.org/10.1002/qute.201900141},
  number   = {11},
  pages    = {1900141},
  url      = {https://onlinelibrary.wiley.com/doi/abs/10.1002/qute.201900141},
  volume   = {3},
  journal  = {Advanced Quantum Technologies},
  year     = {2020},
}

@Article{Langer2025,
  author  = {Langer, M. and Ruchka, P. and Rahimi, A. and Jakovljevic, S. and Zena, Y. G. and Dhurjati, S. A. and Danilov, A. and Pal, M. and Bassoli, R. and Fitzek, F. H. P. and Schmidt, O. G. and Giessen, H. and Hopfmann, C.},
  title   = {An ultra-compact deterministic source of maximally entangled photon pairs},
  doi     = {10.1063/5.0271023},
  issn    = {2378-0967},
  number  = {6},
  pages   = {066117},
  volume  = {10},
  journal = {APL Photonics},
  month   = {06},
  year    = {2025},
}

@Article{Anwar2022,
  author       = {Anwar, Ali and Perumangatt, Chithrabhanu and Villar, Aitor and Lohrmann, Alexander and Ling, Alexander},
  date         = {2022},
  journaltitle = {Applied Physics Letters},
  title        = {Development of compact entangled photon-pair sources for satellites},
  volume       = {121},
  number       = {22},
  pages        = {220503},
  doi          = {10.1063/5.0109702},
  journal      = {Appl. Phys. Lett.},
  refid        = {Anwar2022},
  year         = {2022},
}

@Article{Bremer2020,
  author  = {Bremer, Lucas and Weber, Ksenia and Fischbach, Sarah and Thiele, Simon and Schmidt, Marco and Kaganskiy, Arsenty and Rodt, Sven and Herkommer, Alois and Sartison, Marc and Portalupi, Simone Luca and Michler, Peter and Giessen, Harald and Reitzenstein, Stephan},
  title   = {Quantum dot single-photon emission coupled into single-mode fibers with 3D printed micro-objectives},
  doi     = {10.1063/5.0014921},
  issn    = {2378-0967},
  number  = {10},
  pages   = {106101},
  url     = {https://doi.org/10.1063/5.0014921},
  volume  = {5},
  journal = {APL Photonics},
  month   = {10},
  year    = {2020},
}

@Article{James2001,
  author    = {James, Daniel F. V. and Kwiat, Paul G. and Munro, William J. and White, Andrew G.},
  title     = {Measurement of qubits},
  doi       = {10.1103/PhysRevA.64.052312},
  issue     = {5},
  pages     = {052312},
  url       = {https://link.aps.org/doi/10.1103/PhysRevA.64.052312},
  volume    = {64},
  journal   = {Phys. Rev. A},
  month     = {10},
  numpages  = {15},
  publisher = {American Physical Society},
  year      = {2001},
}

@Article{Winik2017,
  author    = {Winik, R. and Cogan, D. and Don, Y. and Schwartz, I. and Gantz, L. and Schmidgall, E. R. and Livneh, N. and Rapaport, R. and Buks, E. and Gershoni, D.},
  title     = {On-demand source of maximally entangled photon pairs using the biexciton-exciton radiative cascade},
  doi       = {10.1103/PhysRevB.95.235435},
  issue     = {23},
  pages     = {235435},
  url       = {https://link.aps.org/doi/10.1103/PhysRevB.95.235435},
  volume    = {95},
  journal   = {Phys. Rev. B},
  month     = {6},
  numpages  = {7},
  publisher = {American Physical Society},
  year      = {2017},
}

@Article{Kimble2008,
  author       = {Kimble, H Jeff},
  date         = {2008},
  journaltitle = {Nature},
  title        = {The quantum internet},
  number       = {7198},
  pages        = {1023--1030},
  volume       = {453},
  publisher    = {Nature Publishing Group},
}

@Article{Ball2018,
  author       = {Ball, Philip},
  date         = {2018-09},
  journaltitle = {National Science Review},
  title        = {Jian-Wei Pan: building the quantum internet},
  doi          = {10.1093/nsr/nwy102},
  issn         = {2095-5138},
  number       = {2},
  pages        = {374-376},
  url          = {https://doi.org/10.1093/nsr/nwy102},
  volume       = {6},
}

@Article{Gao2022,
  author       = {Timm Gao and Lucas Rickert and Felix Urban and Jan Große and Nicole Srocka and Sven Rodt and Anna Musiał and Kinga Żołnacz and Paweł Mergo and Kamil Dybka and Wacław Urbańczyk and Grzegorz Sȩk and Sven Burger and Stephan Reitzenstein and Tobias Heindel},
  date         = {2022},
  journaltitle = {Applied Physics Reviews},
  title        = {A quantum key distribution testbed using a plug\&amp;play telecom-wavelength single-photon source},
  doi          = {10.1063/5.0070966},
  issn         = {1931-9401},
  number       = {1},
  volume       = {9},
  publisher    = {AIP Publishing},
}

@Article{Li2023,
  author         = {Li, Rusong and Liu, Fengqi and Lu, Quanyong},
  date           = {2023},
  journaltitle   = {Photonics},
  title          = {Quantum Light Source Based on Semiconductor Quantum Dots: A Review},
  doi            = {10.3390/photonics10060639},
  issn           = {2304-6732},
  number         = {6},
  url            = {https://www.mdpi.com/2304-6732/10/6/639},
  volume         = {10},
  article-number = {639},
}

@Article{Liu2022,
  author       = {Ruiqi Liu and Georgi Gary Rozenman and Neel Kanth Kundu and Daryus Chandra and Debashis De},
  date         = {2022},
  journaltitle = {IET Quantum Communication},
  title        = {Towards the industrialisation of quantum key distribution in communication networks: A short survey},
  doi          = {10.1049/qtc2.12044},
  issn         = {2632-8925},
  number       = {3},
  pages        = {151-163},
  volume       = {3},
  publisher    = {Institution of Engineering and Technology (IET)},
}

@Article{Musial2020,
  author       = {Musiał, Anna and Żołnacz, Kinga and Srocka, Nicole and Kravets, Oleh and Große, Jan and Olszewski, Jacek and Poturaj, Krzysztof and Wójcik, Grzegorz and Mergo, Paweł and Dybka, Kamil and Dyrkacz, Mariusz and Dłubek, Michał and Lauritsen, Kristian and Bülter, Andreas and Schneider, Philipp-Immanuel and Zschiedrich, Lin and Burger, Sven and Rodt, Sven and Urbańczyk, Wacław and Sęk, Grzegorz and Reitzenstein, Stephan},
  date         = {2020},
  journaltitle = {Advanced Quantum Technologies},
  title        = {Plug\&Play Fiber-Coupled 73 {kHz} Single-Photon Source Operating in the Telecom O-Band},
  doi          = {https://doi.org/10.1002/qute.202000018},
  number       = {6},
  pages        = {2000018},
  url          = {https://onlinelibrary.wiley.com/doi/abs/10.1002/qute.202000018},
  volume       = {3},
  journal      = {Advanced Quantum Technologies},
}

@Article{Northeast2021,
  author       = {Northeast, David B. and Dalacu, Dan and Weber, John F. and Phoenix, Jason and Lapointe, Jean and Aers, Geof C. and Poole, Philip J. and Williams, Robin L.},
  date         = {2021},
  journaltitle = {Scientific Reports},
  title        = {Optical fibre-based single photon source using InAsP quantum dot nanowires and gradient-index lens collection},
  doi          = {10.1038/s41598-021-02287-y},
  issn         = {2045-2322},
  number       = {1},
  pages        = {22878},
  url          = {https://doi.org/10.1038/s41598-021-02287-y},
  volume       = {11},
  refid        = {Northeast2021},
}

@Article{Rickert2025,
  author       = {Rickert, Lucas and Żołnacz, Kinga and Vajner, Daniel A. and von Helversen, Martin and Rodt, Sven and Reitzenstein, Stephan and Liu, Hanqing and Li, Shulun and Ni, Haiqiao and Wyborski, Paweł and Sęk, Grzegorz and Musiał, Anna and Niu, Zhichuan and Heindel, Tobias},
  date         = {2025-01},
  journaltitle = {Nanophotonics},
  title        = {A fiber-pigtailed quantum dot device generating indistinguishable photons at GHz clock-rates},
  doi          = {10.1515/nanoph-2024-0519},
  issn         = {2192-8614},
  url          = {http://dx.doi.org/10.1515/nanoph-2024-0519},
  publisher    = {Walter de Gruyter GmbH},
}

@Article{Snijders2018,
  author       = {Snijders, H. and Frey, J. A. and Norman, J. and Post, V. P. and Gossard, A. C. and Bowers, J. E. and van Exter, M. P. and L\"offler, W. and Bouwmeester, D.},
  date         = {2018-03},
  journaltitle = {Phys. Rev. Appl.},
  title        = {Fiber-Coupled Cavity-QED Source of Identical Single Photons},
  doi          = {10.1103/PhysRevApplied.9.031002},
  issue        = {3},
  pages        = {031002},
  url          = {https://link.aps.org/doi/10.1103/PhysRevApplied.9.031002},
  volume       = {9},
  numpages     = {6},
  publisher    = {American Physical Society},
}

@Article{Wang2026,
  author         = {Wang, Jing and Li, Peng and Sun, Luyi and Wang, Pengcheng and Li, Nachuan and Zhang, Xiao-Tian and Gong, Yan-Xiao and Liu, Hua-Ying and Zhu, Shi-Ning and Xie, Zhenda},
  date           = {2026},
  journaltitle   = {Photonics},
  title          = {A Compact and Robust Polarization-Entangled Photon Source Towards Application in Mobile Platforms},
  doi            = {10.3390/photonics13020184},
  issn           = {2304-6732},
  number         = {2},
  url            = {https://www.mdpi.com/2304-6732/13/2/184},
  volume         = {13},
  article-number = {184},
}

@Article{Wang2026a,
  author       = {Wang, Jipeng and Hanel, Joscha and Jiang, Zenghui and Joos, Raphael and Jetter, Michael and Rugeramigabo, Eddy Patrick and Portalupi, Simone Luca and Michler, Peter and Cao, Xiao-Yu and Yin, Hua-Lei and Shan, Lei and Yang, Jingzhong and Zopf, Michael and Ding, Fei},
  date         = {2026},
  journaltitle = {Light: Science \& Applications},
  title        = {Time-bin encoded quantum key distribution over 120 km with a telecom quantum dot source},
  doi          = {10.1038/s41377-026-02205-9},
  issn         = {2047-7538},
  number       = {1},
  pages        = {126},
  url          = {https://doi.org/10.1038/s41377-026-02205-9},
  volume       = {15},
  refid        = {Wang2026},
}

@Article{Yang2024,
  author       = {Yang, Jingzhong and Jiang, Zenghui and Benthin, Frederik and Hanel, Joscha and Fandrich, Tom and Joos, Raphael and Bauer, Stephanie and Kolatschek, Sascha and Hreibi, Ali and Rugeramigabo, Eddy Patrick and Jetter, Michael and Portalupi, Simone Luca and Zopf, Michael and Michler, Peter and Kück, Stefan and Ding, Fei},
  date         = {2024},
  journaltitle = {Light: Science \& Applications},
  title        = {High-rate intercity quantum key distribution with a semiconductor single-photon source},
  doi          = {10.1038/s41377-024-01488-0},
  issn         = {2047-7538},
  number       = {1},
  pages        = {150},
  url          = {https://doi.org/10.1038/s41377-024-01488-0},
  volume       = {13},
  refid        = {Yang2024},
}

@Article{Eisaman2011,
  author       = {Eisaman, M. D. and Fan, J. and Migdall, A. and Polyakov, S. V.},
  date         = {2011-07},
  journaltitle = {Review of Scientific Instruments},
  title        = {Invited Review Article: Single-photon sources and detectors},
  doi          = {10.1063/1.3610677},
  issn         = {0034-6748},
  number       = {7},
  pages        = {071101},
  url          = {https://doi.org/10.1063/1.3610677},
  volume       = {82},
}

@Article{Keil2017,
  author       = {Keil, Robert and Zopf, Michael and Chen, Yan and H{\"o}fer, Bianca and Zhang, Jiaxiang and Ding, Fei and Schmidt, Oliver G},
  date         = {2017},
  journaltitle = {Nat. Commun.},
  title        = {Solid-state ensemble of highly entangled photon sources at rubidium atomic transitions},
  number       = {1},
  pages        = {1--8},
  volume       = {8},
  publisher    = {Nature Publishing Group},
}

@Book{Michler2003,
  date      = {2003},
  title     = {Single Quantum Dots},
  editor    = {Peter Michler},
  isbn      = {9783540391807},
  location  = {Berlin [u.a.]},
  number    = {90},
  publisher = {Springer},
  series    = {Topics in Applied Physics},
  subtitle  = {Fundamentals, Applications, and New Concepts},
  url       = {http://dx.doi.org/10.1007/b13751},
  ppn_gvk   = {826349838},
}

@Article{Sibson2017,
  author       = {Philip Sibson and Jake E. Kennard and Stasja Stanisic and Chris Erven and Jeremy L. O’Brien and Mark G. Thompson},
  title        = {Integrated silicon photonics for high-speed quantum key distribution},
  issn         = {2334-2536},
  number       = {2},
  pages        = {172},
  volume       = {4},
  date         = {2017},
  doi          = {10.1364/optica.4.000172},
  journaltitle = {Optica},
  publisher    = {Optica Publishing Group},
}
\end{document}